\documentclass[12pt]{article}
\usepackage[margin=1in]{geometry}
\usepackage[numbers]{natbib}
\usepackage{amsmath,amssymb,authblk,bm,enumerate,graphicx,rotating}
\usepackage[dvipsnames,table,dvipsnames*, svgnames*, hyperref]{xcolor}

\DeclareMathOperator*{\esssup}{ess\,sup}
\def\E{\mathrm{E}}

\def\var{\mathrm{var}\,}

\newtheorem{lemma}{Lemma}
\newtheorem{theorem}{Theorem}
\newtheorem{assumption}{Assumption}

\title{Optimal detection of weak positive latent dependence between two sequences of multiple tests}
\author[1]{Sihai Dave Zhao}
\affil[1]{Department of Statistics, University of Illinois at Urbana-Champaign}
\author[2]{T. Tony Cai}
\affil[2]{Department of Statistics, The Wharton School, University of Pennsylvania}
\author[3]{Hongzhe Li}
\affil[3]{Department of Biostatistics and Epidemiology, Perelman School of Medicine, University of Pennsylvania}
\date{}

\begin{document}
\maketitle

\begin{abstract}
  It is frequently of interest to jointly analyze two paired sequences of multiple tests. This paper studies the problem of detecting whether there are more pairs of tests that are significant in both sequences than would be expected by chance. The asymptotic detection boundary is derived in terms of parameters such as the sparsity of non-null cases in each sequence, the effect sizes of the signals, and the magnitude of the dependence between the two sequences. A new test for detecting weak dependence is also proposed, shown to be asymptotically adaptively optimal, studied in simulations, and applied to study genetic pleiotropy in 10 pediatric autoimmune diseases.
\end{abstract}

\section{\label{sec:intro}Introduction}
\subsection{Overview}
Joint analysis of two paired sequences of multiple tests, each arising from a separate independent study, arises in many applications. It has been particularly motivated by genomics research, where it is natural to investigate similarities in how genomic features, such as genes or genetic markers, behave across studies. For example, recent interest has focused on features that may be significant in both of two sequences of multiple tests. In differential gene expression experiments, enrichment analysis \citep{huang2009bioinformatics} is often used to test whether two experiments share more significantly differentially expressed genes than would be expected by chance. In the integration of an expression quantitative trait loci study and a genome-wide association study, the goal is frequently to detect and identify genetic variants that are associated with both gene expression and disease \citep{he2013sherlock,nicolae2010trait}. Replicability analysis \citep{heller2014deciding,heller2014repfdr,heller2014replicability} aims to discover significant findings that have been replicated across genomic studies. Finally, studies of genetic pleiotropy investigate whether the same genetic variants may be simultaneously associated with different traits \citep{brown2016transethnic,chung2014gpa,cross2013genetic,cross2013identification,guo2016optimal,lee2012estimating}.

These examples broadly fall into two categories of questions: the detection of whether there exist features that are significant in both of two studies, and the identification of those simultaneously significant features. This paper focuses on the detection problem; the identification problem is studied elsewhere \citep{chung2014gpa,heller2014replicability,phillips2014testing,zhao2017false}. Specifically, let $I_{kj}$ be unobserved latent indicators of whether the $j$th test, $j=1,\ldots,p,$ is truly non-null in the $k$th study, $k=1,2$. Let $T_{kj}$ be the corresponding test statistic such that
\begin{equation}
  \label{eq:mixture}
  T_{kj}\mid I_{kj}=0\sim F^0_k,
  \quad
  T_{kj}\mid I_{kj}=1\sim F^1_k,
  \quad
  I_{kj}\sim Ber(\pi_k),
  \quad
  k=1,2,
\end{equation}
where the $\pi_k$ quantify the proportion of non-null tests in each study. The $F^0_k$ and $F^1_k$ can be viewed as mixtures of possibly different null and non-null distributions for different $j$. For each $k$, model~\eqref{eq:mixture} corresponds to a two-group mixture model for $T_{kj}$, which is common in the literature \citep{donoho2004higher,efron2010large,genovese2002operating,storey2003statistical,sun2007oracle}. It will be assumed that the $T_{kj}$ are two-tailed test statistics and are thus stochastically larger when $I_{kj}=1$. Because the two sequences of tests arise from different studies, which typically are conducted on independent samples, it is assumed that $T_{1j}$ and $T_{2j}$ are independent conditional on the latent indicators $I_{1j}$ and $I_{2j}$.

The goal of this paper is to test whether there are more features $j$ that are significant in both studies than would be expected by chance. Formally, if $\Pr(I_{1j}=1,I_{2j}=1)=\epsilon$, the goal is to test
\begin{equation}
  \label{eq:test}
  H_0:\epsilon=\pi_1\pi_2
  \quad
  vs.
  \quad
  H_A:\epsilon>\pi_1\pi_2.
\end{equation}
This is motivated by a study of genetic pleiotropy in 10 pediatric autoimmune diseases conducted by Hakonarson and colleagues at the Children's Hospital of Pennsylvania \citep{li2015meta,li2015genetic}. More details about the data can be found in Section~\ref{sec:application}. Testing~\eqref{eq:test} using genome-wide association study summary statistics from a pair of diseases can assess whether the two conditions have some degree of shared genetic architecture, which can lead to a better understanding of their etiologies.

Several features make testing~\eqref{eq:test} difficult for existing methods. First, the $I_{kj}$ are not directly observed. Second, in genomics applications, non-null features are typically rare and have weak effect sizes. For example, only a relatively small proportion of the human genome is expected to be associated with a given phenotype, and then only weakly so. Finally, positive dependence between $I_{1j}$ and $I_{2j}$ can be very weak when it exists, because cross-study heterogeneity makes it unlikely that more than a handful of features will be simultaneously non-null in both of two independently conducted genomics studies, even if the studies are closely related.

This paper proposes a new test for~\eqref{eq:test} under these challenging conditions. The proposed test statistic is shown to be asymptotically adaptively optimal, so that it performs as well as the optimal likelihood ratio test statistic but without needing to specify parameter values under $H_0$ and $H_A$. In fact the proposed test is entirely nonparametric, so neither $F^0_k$ nor $F^1_k$ needs to be known. It is also computationally efficient to implement and can be computed for 10 million pairs of tests in under one minute. It is available in the \textsf{R} package \texttt{ssa}.

\subsection{Related work}
Because model~\eqref{eq:mixture} assumes that $T_{1j}$ and $T_{2j}$ are independent conditional on $I_{1j}$ and $I_{2j}$, testing \eqref{eq:test} is equivalent to testing for independence between $T_{1j}$ and $T_{2j}$. Classical methods are based on goodness-of-fit tests comparing the empirical bivariate distribution of $(T_{1j},T_{2j})$ to the product of the marginal empirical distributions. Variations include Cramer-von-Mises-, Anderson-Darling-, and Kolmogorov-Smirnov-type tests \citep{hoeffding1948non,scaillet2005kolmogorov,thas2004nonparametric}. A number of methods for detecting positive quadrant dependence have also been studied in the actuarial sciences \citep{ledwina2014validation}. Independence testing has seen renewed interest in the statistical literature, where the focus is on detecting arbitrary types of dependence \citep{reshef2011detecting,szekely2009brownian}; see in particular \citet{heller2016consistent}. In contrast, this paper is concerned with detecting a particular form of dependence between $T_{1j}$ and $T_{2j}$, induced by the weak positive latent dependence between $I_{1j}$ and $I_{2j}$. It appears that this type of dependence has not yet been specifically considered, and existing methods may be suboptimal. Furthermore, the fundamental limits of detection have not been studied.

Testing \eqref{eq:test} can also be seen as an extension of the single-sequence signal detection problem. There, given test statistics $T_{kj}$ from a single study $k$, the goal is to determine whether there are any non-null signals: $H_0:\Pr(I_{kj}=1)=0$ vs. $H_A:\Pr(I_{kj}=1)>0$. The fundamental limits of detection for this problem have been derived, and asymptotically adaptively optimal tests have also been developed \citep{arias2017distribution,cai2011optimal,cai2014optimal,delaigle2011robustness,donoho2004higher,ingster1997some,ingster2002adaptiveI,ingster2002adaptiveII,jager2007goodness}. Special attention has been paid to the setting where $\pi_k$ is very close to zero and $F^1_k$ is not too different from $F^0_k$. As previously noted, this rare and weak signal setting is also the focus of this paper. However, results for the single sequence problem do not apply to testing~\eqref{eq:test}.

Several additional methods for testing \eqref{eq:test} have been developed in the genomics literature. A popular approach is to estimate the $I_{kj}$, by thresholding the $T_{kj}$, and then to test for dependence using the estimated $I_{kj}$ \citep{huang2009bioinformatics,rivals2007enrichment}. However, it is unclear how the thresholds on $T_{kj}$ should be chosen. Alternatively, the GPA method \citep{chung2014gpa} fits the $(T_{1j},T_{2j})$ to a four-group mixture model, each group corresponding to one of the four possible values of the tuple $(I_{1j},I_{2j})$, and uses a generalized likelihood ratio test for~\eqref{eq:test}. However, GPA imposes parametric assumptions on $F^0_k$ and $F^1_k$. In addition, theoretical results from the single-sequence detection problem suggests that generalized likelihood ratio tests will have poor asymptotic properties when non-null $T_{kj}$ are rare and weak \citep{hartigan1985failure}. Recently, \citet{zhao2017sparse} proposed a simple test for~\eqref{eq:test} and studied its asymptotic properties. However, their theoretical results require distributional assumptions on the $T_{kj}$, and their test is only asymptotically optimal under specialized conditions.

The rest of the paper is organized as follows. Section~\ref{sec:method} introduces the proposed test statistic and Section~\ref{sec:theory} studies its asymptotic adaptive optimality. Section~\ref{sec:sims} presents the results of simulation studies and the pediatric autoimmune disease analysis. The paper concludes with a discussion in Section~\ref{sec:discussion}. Additional simulation and data analysis results, and all proofs, can be found in the Appendix.

\section{\label{sec:method}Proposed method}
\subsection{\label{sec:teststat}Test statistic}
Because testing~\eqref{eq:test} is equivalent to detecting dependence between $T_{1j}$ and $T_{2j}$, let $\hat{S}_{12}(t_1,t_2)$ and $\hat{S}_k(t_k)$ denote the empirical bivariate and marginal survival functions, respectively:
\[
\hat{S}_{12}(t_1,t_2)=p^{-1}\sum_{j=1}^pI(T_{1j}\geq t_1,T_{2j}\geq t_2),
\quad
\hat{S}_k(t_k)=p^{-1}\sum_{j=1}^pI(T_{kj}\geq t_k),\quad k=1,2.
\]
The proposed test statistic is
\begin{equation}
  \label{eq:Dhat}
  \widehat{\mathcal{D}}
  =
  \sup_{(t_1,t_2)\in\mathcal{S}}p^{1/2}\frac{\vert\hat{S}_{12}(t_1,t_2)-\hat{S}_1(t_1)\hat{S}_2(t_2)\vert}{\{\hat{S}_1(t_1)\hat{S}_2(t_2)-\hat{S}_1^2(t_1)\hat{S}_2^2(t_2)\}^{1/2}},
\end{equation}
where the set $\mathcal{S}$ is defined as
\begin{equation}
  \label{eq:S}
  \mathcal{S}=[T_{1(1)},T_{1(p)}]\times[T_{2(1)},T_{2(p)}]\setminus\{(T_{1(1)},T_{2(1)})\},
\end{equation}
and $T_{k(j)}$ is the $j$th order statistics of the $T_{kj}$. This is the supremum version of an Anderson-Darling-type goodness-of-fit test for independence between $T_{1j}$ and $T_{2j}$, and is motivated by the higher criticism statistic of \citet{donoho2004higher} for signal detection in a single sequence of multiple tests. Properties of an oracle version of the statistic $\widehat{\mathcal{D}}$, where the $\hat{S}_k$ are replaced by the true marginal survival functions, have been previously studied \citep{einmahl1996extension,einmahl1985bounds}, but not in the present context of weak latent dependency detection. One advantage of $\widehat{\mathcal{D}}$ is that it makes no assumptions about the distributions $F^0_k$ and $F^1_k$.

\begin{figure}
  \centering
  \includegraphics[scale=0.5]{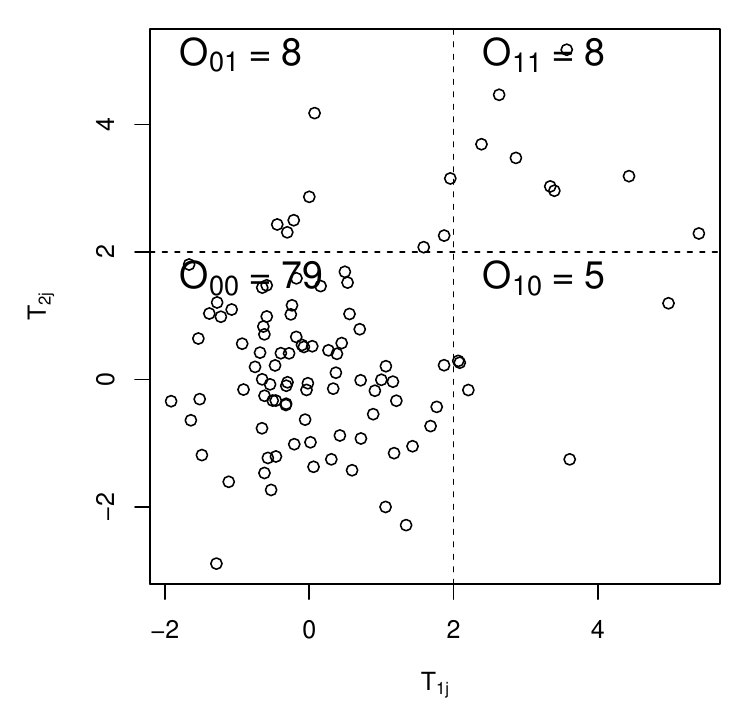}
  \caption{\label{fig:2x2}The $2\times2$ table induced in $(T_{1j},T_{2j}),j=1,\ldots,100$, generated according to~\eqref{eq:data_example}, by the tuple $(t_1,t_2)=(2,2)$. The cell counts are denoted by $O_{lm}$, $l,m=0,1$.}
\end{figure}

To better understand its properties, first consider the numerator $\hat{S}_{12}-\hat{S}_1\hat{S}_2$. This is a natural way to test for dependence between $T_{1j}$ and $T_{2j}$ and thus~\eqref{eq:test}, but there is a useful alternative interpretation. Figure~\ref{fig:2x2} is a scatterplot of 100 realizations from the following data-generating mechanism:
\begin{equation}
  \label{eq:data_example}
  \begin{aligned}
    &T_{kj}\mid I_{kj}=0\sim \mathcal{N}(0,1),\quad T_{kj}\mid I_{kj}=1\sim \mathcal{N}(3,1),k=1,2,\\
    &\Pr(I_{1j}=1,I_{2j}=1)=0.1,
    \quad
    \Pr(I_{1j}=1,I_{2j}=0)=0.05,
    \quad
    \Pr(I_{1j}=0,I_{2j}=1)=0.05,\\
    &
    \Pr(I_{1j}=0,I_{2j}=0)=0.8.
  \end{aligned}
\end{equation}
The figure illustrates that any tuple $(t_1,t_2)$ divides the observed data into a $2\times2$ contingency table. \citet{blum1961distribution} recognized that the numerator is closely related to testing for independence using the cell counts of the $2\times2$ table induced by $(t_1,t_2)$. Later, \citet{thas2004nonparametric} and most recently \citet{heller2016consistent} extended this idea to $m\times m$ tables for $m\geq2$, which \citet{heller2016consistent} showed can have greater power.

Next consider the supremum in $\widehat{\mathcal{D}}$. It is difficult to know {\it a priori} which tuple $(t_1,t_2)$ will induce the $2\times2$ table that gives the largest test statistic. The optimal $(t_1,t_2)$ depends on the distributions $F^0_k$ and $F^1_k$, the proportions $\pi_k$, and the degree of dependence $\epsilon$. Thus $\widehat{\mathcal{D}}$ takes the supremum over all possible $(t_1,t_2)$, allowing it to adapt to any combination of these unknown parameters. Instead of the supremum, \citet{thas2004nonparametric} proposed a statistic that integrates over all tuples; their statistic turns out to be closely related to to summing the Pearson chi-square test statistics calculated from each $2\times2$ table induced by each of the observed tuples $(T_{1j},T_{2j})$. \citet{heller2016consistent} proposed several procedures that either sum or take the maximum over statistics arising from all possible $m\times m$ tables, then combines these statistics across multiple choices for $m$.

\begin{figure}
  \centering
  \includegraphics[scale=0.4]{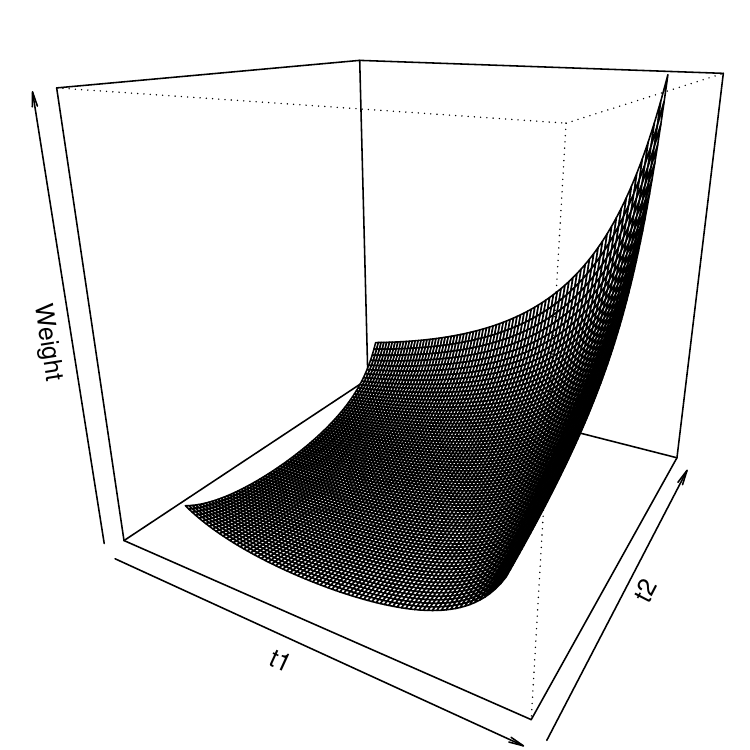}
  \caption{\label{fig:Dweight}Plot of the weight function $\{S_1(t_1)S_2(t_2)-S_1^2(t_1)S_2^2(t_2)\}^{-1/2}$ for $S_k(x)=1-x$.}
\end{figure}

Finally, consider the denominator of $\widehat{\mathcal{D}}$. It is a natural standardizing weight in that it is the variance of $\hat{S}_{12}$ under the independence null hypothesis of~\eqref{eq:test}. Furthermore, it is the reason why $\widehat{\mathcal{D}}$ can have power for detecting even weak dependence. Figure~\ref{fig:Dweight} plots the inverse of the denominator when the marginal survival functions are known and equal to $S_k(x)=1-x$. It is largest for large $t_1$ and $t_2$, which corresponds to the upper right-hand quadrant of Figure~\ref{fig:2x2}. This implies that $\widehat{\mathcal{D}}$ can be large even when only a few points are observed in this quadrant, which will be the case when the $T_{kj}$ are stochastically larger when $I_{kj}=1$ but only weakly dependent. Other denominators are also possible but may not be optimal for detecting weak positive latent dependence; see Section~\ref{sec:alt}.

\subsection{\label{sec:inf}Inference}
When the test statistics $T_{kj}$ are independent across $j$, \citet{einmahl1996extension} showed that the oracle statistic
\begin{equation}
  \label{eq:oracle}
  \mathcal{D}
  =
  \sup_{-\infty<t_1,t_2<\infty}p^{1/2}\frac{\vert\hat{S}_{12}(t_1,t_2)-S_1(t_1)S_2(t_2)\vert}{\{S_1(t_1)S_2(t_2)-S_1^2(t_1)S_2^2(t_2)\}^{1/2}},
\end{equation}
where the marginal survival functions are known, satisfies
\[
\Pr_{H_0}\{(\ln p)^{-1/2}\mathcal{D}>x\}\rightarrow1-\exp(-x^2)
\]
under the null hypothesis $H_0$ of independence between $I_{1j}$ and $I_{2j}$. However, this oracle result may not be applicable to the proposed $\widehat{\mathcal{D}}$~\eqref{eq:Dhat}. Furthermore, the convergence rates of these types of extreme values statistics are usually too slow to be useful \citep{barnett2014analytical,cai2011optimal,donoho2004higher}.

Instead, this paper considers a simple permutation procedure to provide $p$-values. Fixing the indices of $T_{2j}$, randomly permute the indices of $T_{1j}$ to induce independence between the two sequences of tests. Let $\widehat{\mathcal{D}}^{(l)}$ be the proposed statistic~\eqref{eq:Dhat} calculated after the $l$th permutation. Then the $p$-value after $B$ permutations is $\{1+\sum_{l=1}^BI(\widehat{\mathcal{D}}^{(l)}\geq\widehat{\mathcal{D}})\}/(B+1)$ \citep{lehmann2005testing}.
Even large numbers of permutations are feasible because $\widehat{\mathcal{D}}$ can be computed very quickly, as described below in Section~\ref{sec:implementation}.

In many genomic applications, the $T_{kj}$ are likely to be dependent across $j$. For example, if each $T_{kj}$ is the test statistic for association between genetic variant $j$ and phenotype $k$, the $T_{kj}$ will be correlated across $j$ due to linkage disequilibrium. Interestingly, simulations with real genotype data in Section~\ref{sec:dep} indicate that using the random permutation $p$-value is still able to maintain type I error. 

\subsection{\label{sec:implementation}Implementation}
A simple algorithm for calculating $\widehat{\mathcal{D}}$ requires $O(p^2)$ operations: the $T_{1j}$ and $T_{2j}$ are first sorted using quicksort, which on average requires $O(p\ln p)$ operations and at most requires $O(p^2)$. Next, the algorithm iterates from the largest to the smallest order statistics $T_{1(l)}$, where for each $l$ it iterates from the largest to the smallest $T_{2(m)}$ in order to calculate
\[
D_{lm}=p^{1/2}\frac{\vert\hat{S}_{12}(T_{1(l)},T_{2(m)})-\hat{S}_1(T_{1(l)})\hat{S}_2(T_{2(m)})\vert}{\{\hat{S}_1(T_{1(l)})\hat{S}_2(T_{2(m)})-\hat{S}_1(T_{1(l)})^2\hat{S}_2(T_{(m)})^2\}^{1/2}}
\]
for all $l,m=1,\ldots,p$. Finally, $\widehat{\mathcal{D}}=\max_{lm}D_{lm}$. This algorithm has been implemented in $C$ in the \textsf{R} package \texttt{ssa}.

An additional computational shortcut can be implemented. Because the $T_{kj}$ are stochastically larger when $I_{kj}=1$, the largest $D_{lm}$ is likely to be found when $l$ and $m$ are large. Therefore the algorithm only needs to iterate over $T_{1(p-m_1+1)},\ldots,T_{1(p)}$ and $T_{2(p-m_2+1)},\ldots,T_{2(p)}$, where $m_1$ and $m_2$ can be close to $p$. Even if the true maximum $D_{lm}$ is not attained for these test statistics, the largest $D_{ij}$ in this restricted region may still be large enough to reject the null hypothesis. This truncated calculation should at worst provide a conservative test, and $m_1$ and $m_2$ can be set as large as computationally feasible. As an example, this algorithm can calculate $\widehat{\mathcal{D}}$ for $p=10^7$ and $m_1=m_2=10^4$ in 29 seconds on a laptop with a 2.5 GHz Intel Core i5 processor with 8 GB RAM.

\section{\label{sec:theory}Theoretical justification}

\subsection{\label{sec:assumptions}Assumptions}
As introduced in Section~\ref{sec:intro}, for any feature $j$ the observed $T_{1j}$ and $T_{2j}$ are assumed to follow model~\eqref{eq:mixture}. Because they are derived from two different studies, they will be independent conditional on the $I_{kj}$. They are also assumed to be two-tailed test statistics and thus stochastically larger when $I_{kj}=1$ than when $I_{kj}=0$.

\begin{assumption} 
  \label{a:sto_ord}
  For $k=1,2$, $F^1_k(t)\leq F^0_k(t)$ for all $t$.
\end{assumption}

The dependency detection problem~\eqref{eq:test} and the proposed test statistic $\widehat{\mathcal{D}}$~\eqref{eq:Dhat} will be studied under the asymptotic testing framework \citep{lehmann2005testing}, where the asymptotics apply to the total number of tests $p$. This is meaningful because in practice $p$ can be very large, such as in applications to genome-wide association studies. If the parameters $\epsilon$, $\pi_k$, $F^0_k$, and $F^1_k$, $k=1,2$ were fixed with $p$, any reasonable test would be able to distinguish $H_0$ from $H_A$. Instead, the parameters will calibrated to vary with $p$. This allows for a more meaningful comparison between possible testing procedures, and in addition formalizes the setting of weak positive latent dependence and rare and weak signals, described in Section~\ref{sec:intro}.

Specifically, $\epsilon$ and $\pi_k$ will be calibrated to approach 0, which models weak dependence and rare signals:
\begin{equation}
  \label{eq:calibrate}
  \begin{array}{ll}
    \pi_k=p^{-\beta_k},&1/2\leq\beta_k\leq1,k=1,2,\\
    \epsilon=\pi_1\pi_2+p^{-\beta},&1/2<\beta<1,(\beta_1\vee\beta_2)\leq\beta.
  \end{array}
\end{equation}
In genomics problems, typically very few of the $T_{kj}$ are non-null, which is reflected in the regime $1/2\leq\beta_k$ \citep{donoho2004higher,cai2011optimal,cai2014optimal}. Analogously, this paper models weak dependence by letting $\beta>1/2$. The additional restriction $\beta\geq\beta_k$ ensures that $\epsilon\leq(\pi_1\wedge\pi_2)$.

Given \eqref{eq:calibrate}, $F^1_k$ must be calibrated to separate from $F^0_k$, otherwise testing~\eqref{eq:test} would be very difficult. This divergence will be expressed in terms of the likelihood ratio between the two distributions. Because no parametric assumptions are made on $F^1_k$ and $F^0_k$, the exact form of this calibration is fairly abstract. Let $f^1_k$ and $f^0_k$ be the corresponding density functions and let $x\vee y$ denote $\max(x,y)$.

\begin{assumption}
  \label{a:tails}
  There exist measurable functions $\alpha^-_k,\alpha^+_k:\mathbb{R}\rightarrow\mathbb{R}$ such that $\alpha_k(a)=\alpha^-_k(a)\vee\alpha^+_k(a)>0$ on a set of positive Lebesgue measure and that for $k=1,2$, the log-likelihood ratios $\ell_k=\ln(f^1_k/f^0_k)$ satisfy
  \[
  \lim_{p\rightarrow\infty}\frac{\ell_k\{(F^0_k)^{-1}(p^{-a})\}}{\ln p}=\alpha^-_k(a),
  \quad  
  \lim_{p\rightarrow\infty}\frac{\ell_k\{(F^0_k)^{-1}(1-p^{-a})\}}{\ln p}=\alpha^+_k(a),
  \]
  uniformly in $a\geq\log_p2$.
\end{assumption}

Assumption~\ref{a:tails} guarantees the existence of limiting functions $\alpha^-_k$ and $\alpha^+_k$ that characterize the likelihood ratios at small and large values, specifically $p^{-a}$ and $1-p^{-a}$. The assumption essentially calibrates the likelihood ratios to grow only polynomially in $p$, which models weak signals. Restricting $a\geq\log_p2$ is necessary because otherwise the $\alpha$ functions would simply be reparametrizations of each other. Since $p^{-\log_p2}=1-p^{-\log_p2}=0.5$,  $p^{-a}$ and $1-p^{-a}$ correspond to numbers smaller and larger than the median of $F^0_k$, respectively. The value of separately characterizing the likelihood ratios on the left and right sides of the null median will become clear in the theoretical results in Section~\ref{sec:properties}.

Assumption~\ref{a:tails} was used in \citet{cai2014optimal} in their study of the single-sequence detection problem and generalizes similar assumptions made in previous work. For example, suppose $F^0_k\equiv \mathcal{N}(0,1)$. Then $\Phi\{-(2a\ln p)^{1/2}\}\approx p^{-a}$ as long as $2a\ln p$ is sufficiently large, which is guaranteed by the condition $a\geq\log_p2$. Therefore the $(p^{-a})$th quantile of $F^0_k$ is $-(2a\ln p)^{-1/2}$, and by similar reasoning the $(1-p^{-a})$th quantile is $(2a\ln p)^{1/2}$. The setting of $F^1_k\equiv \mathcal{N}\{(2r_k\ln p)^{1/2},1\}$, a popular model for weak signals \citep{cai2011optimal,donoho2004higher,ingster1997some,ingster2002adaptiveI,ingster2002adaptiveII}, can be shown to correspond to
\begin{equation}
  \label{eq:alpha_normal}
  \alpha^-_k(a)=-2(ar_k)^{1/2}-r_k,
  \quad
  \alpha^+_k(a)=2(ar_k)^{1/2}-r_k
\end{equation}
in the notation of Assumption~\ref{a:tails}.

Finally, for the purpose of studying the asymptotic properties of $\widehat{\mathcal{D}}$, it will be assumed that in each sequence of tests, the test statistics are mutually independent. This is a simplification, but for dependent tests the asymptotic theory of these types of detection problems is still under development for arbitrary correlation structures \citep{arias2011global,barnett2017generalized,hall2010innovated,mukherjee2015hypothesis}. In contrast, the theoretical properties when tests are independent are well understood, at least for single-sequence problems \citep{cai2011optimal,cai2014optimal,donoho2004higher}. To facilitate comparison with these established results, this paper assumes that $T_{kj}$ and $T_{kj'}$ are independent for $j\ne j'$, and leaves consideration of dependent tests for future work.

\subsection{\label{sec:properties}Asymptotic properties}
For the proposed $\widehat{\mathcal{D}}$~\eqref{eq:Dhat}, consider the test
\begin{equation}
  \label{eq:adaptive_test}
  \mbox{reject $H_0$ of~\eqref{eq:test} if }\widehat{\mathcal{D}}>\ln p(\ln\ln p)^2+3(\ln\ln p)^2.
\end{equation}
The critical value $\ln p(\ln\ln p)^2+3(\ln\ln p)^2$ is chosen such that test~\eqref{eq:adaptive_test} can achieve type I and type II errors that sum to zero as $p\rightarrow\infty$; this will be shown below. Furthermore, it will also be shown that test~\eqref{eq:adaptive_test} is in a certain sense asymptotically optimal among all possible tests. These results support the use of the proposed $\hat{\mathcal{D}}$ for detecting weak positive latent dependence.

 Theorem~\ref{thm:adaptive} characterizes a region of the parameter space where test~\eqref{eq:adaptive_test} will be successful. This region can be expressed in terms of $\beta_k$ and $\beta$ from calibration~\eqref{eq:calibrate} and $\alpha^-_k$ and $\alpha^+_k$ from Assumption~\ref{a:tails}.

\begin{theorem} 
  \label{thm:adaptive}
  Suppose $F^0_k\ne F^1_k,k=1,2$ and define
  \[
  v^-_k(x)=\esssup_{a\geq x}\{\alpha^-_k(a)-a\},
  \quad
  v^+_k(x)=\esssup_{a\geq x}\{\alpha^+_k(a)-a\}.
  \]
  Under calibration~\eqref{eq:calibrate} and Assumption~\ref{a:tails}, the sum of the type I and II errors of~\eqref{eq:adaptive_test} goes to 0 if one of the following is true:
  \begin{align}
    &\sup_{\substack{x_1,x_2>0,\\x_1+x_2<1}}\left(\frac{1}{2}-\beta+\sum_{k=1}^2\left\{(-x_k)\vee v^+_k(x_k)+\frac{x_k\wedge\{\beta_k-v^+_k(x_k)\}}{2}\right]\right)>0,\mbox{ or}\label{eq:Q1}\\
    &\sup_{\substack{x_1,x_2>0,\\x_2<1}}\left[\frac{1}{2}-\beta+(-x_1)\vee v^-_1(x_1)+(-x_2)\vee v^+_2(x_2)+\frac{x_2\wedge\{\beta_2-v^+_2(x_2)\}}{2}\right]>0,\mbox{ or}\label{eq:Q2}\\
    &\sup_{\substack{x_1,x_2>0,\\x_1<1}}\left[\frac{1}{2}-\beta+(-x_1)\vee v^+_1(x_1)+(-x_2)\vee v^-_2(x_2)+\frac{x_1\wedge\{\beta_1-v^+_1(x_1)\}}{2}\right]>0,\mbox{ or}\label{eq:Q3}\\
    &\sup_{x_1,x_2>0}\left\{\frac{1}{2}-\beta+(-x_1)\vee v^-_1+(-x_2)\vee v^-_2+\frac{x_1\wedge\beta_1\wedge x_2\wedge\beta_2}{2}\right\}>0.\label{eq:Q4}
  \end{align}
\end{theorem}

It is also possible to derive the fundamental limits of detecting weak positive latent dependence~\eqref{eq:test}. Theorem~\ref{thm:undetectable} characterizes a region of the parameter space where successful detection is impossible, in the sense that the sum of the type I and II errors of any hypothesis test of~\eqref{eq:test} goes to at least 1 as $p\rightarrow\infty$. It involves the essential supremum, which for a measurable function $f$ and a measure $\mu$ is defined as
\[
\esssup_xf(x)=\inf[a\in\mathbb{R}:\mu\{f(x)>a\}=0].
\]
Here, essential suprema are taken with respect to the Lebesgue measure.

\begin{theorem}
  \label{thm:undetectable}
  Suppose $F^0_k\ne F^1_k,k=1,2$. Under calibration~\eqref{eq:calibrate} and Assumption~\ref{a:tails}, the sum of the type I and II errors of any test goes to at least 1 if each of the following holds:
  \begin{align}
    &1-2\beta+\esssup_{a>0}\{\alpha_k(a)+\alpha_k(a)\wedge\beta_k-a\}<0,\quad k=1,2,\mbox{ and}\label{eq:undetectable1}\\
    &1+\esssup_{a_1,a_2>0}[\{-\beta+\alpha_1(a_1)+\alpha_2(a_2)\}\wedge\nonumber\\
    &\{-2\beta+\alpha_1(a_1)+\alpha_2(a_2)+\alpha_1(a_1)\wedge\beta_1+\alpha_2(a_2)\wedge\beta_2\}-a_1-a_2]<0\label{eq:undetectable2},
  \end{align}
  where $\alpha_k(a)=\alpha^-_k(a)\vee\alpha^+_k(a)$ as defined in Assumption~\ref{a:tails}.
\end{theorem}

When the $T_{kj}$ are stochastically ordered according to Assumption~\ref{a:sto_ord}, it turns out that the union of the two regions defined in Theorems~\ref{thm:adaptive} and~\ref{thm:undetectable}, and the boundary that separates them, constitutes the entire parameter space. In other words, this boundary, called the detection boundary, partitions the parameter space into two regions. In the undetectable region, successful detection is impossible for any test, while in the detectable region, there exists a test, namely~\eqref{eq:adaptive_test}, that can perfectly separate $H_0$ and $H_A$.

\begin{theorem}
  \label{thm:boundary}
  Under Assumption~\ref{a:sto_ord}, the region described by Theorem~\ref{thm:undetectable} is the interior of the complement of the region described by Theorem~\ref{thm:adaptive}. In particular, the detectable region is entirely described by inequality~\eqref{eq:Q1}.
\end{theorem}

The asymptotic optimality of the proposed $\widehat{\mathcal{D}}$ is encapsulated in Theorem~\ref{thm:boundary}. It implies that whenever detection of weak positive latent dependence is possible, \eqref{eq:adaptive_test} already achieves asymptotically zero error. In other words, it can perform as well as the the optimal likelihood ratio test, but has the added benefit that it is entirely data-driven and automatically adapts to the unknown values of $\beta$, $\beta_k$, $F^0_k$, and $F^1_k$ under both $H_0$ and $H_A$.

\begin{figure}
  \centering
  \includegraphics[width=0.48\textwidth]{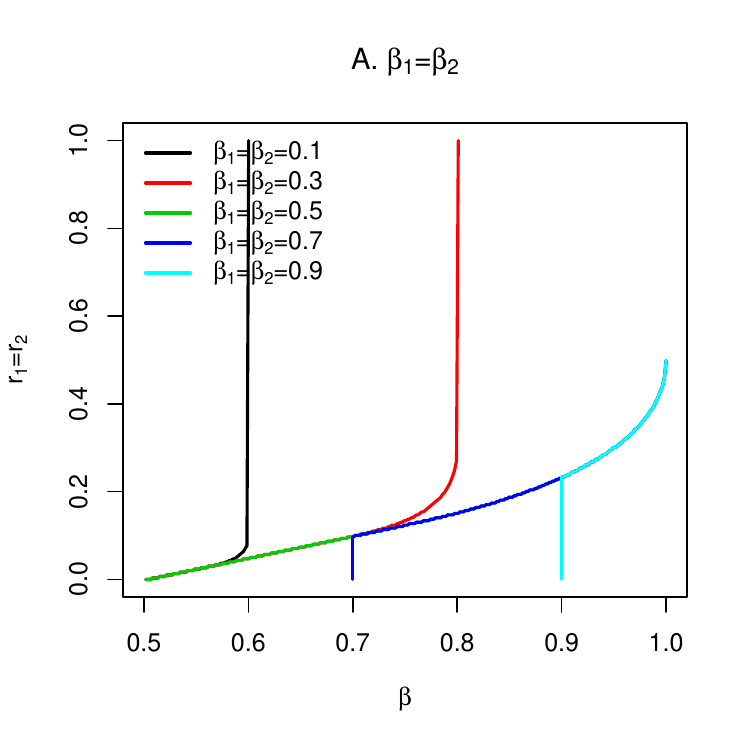}
  \includegraphics[width=0.48\textwidth]{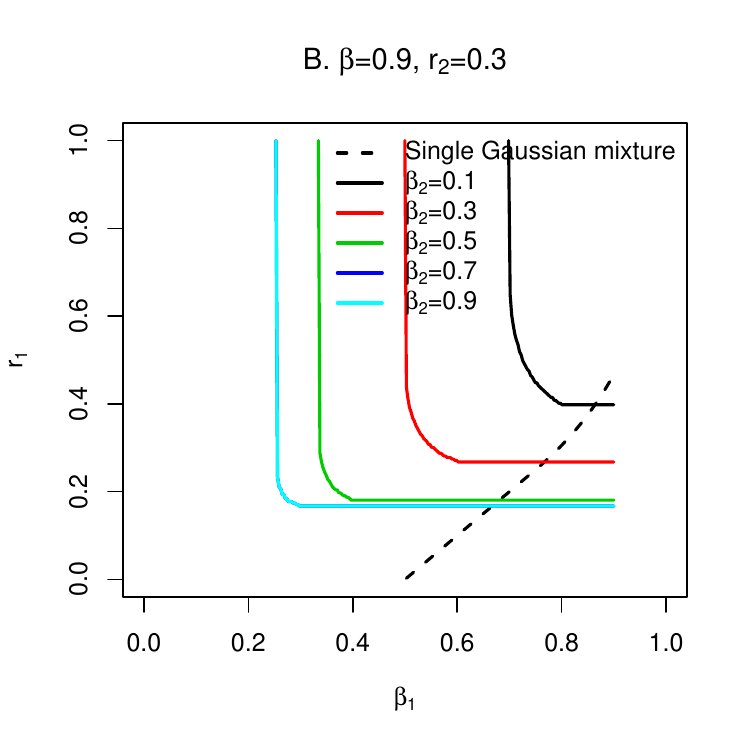}
  \includegraphics[width=0.48\textwidth]{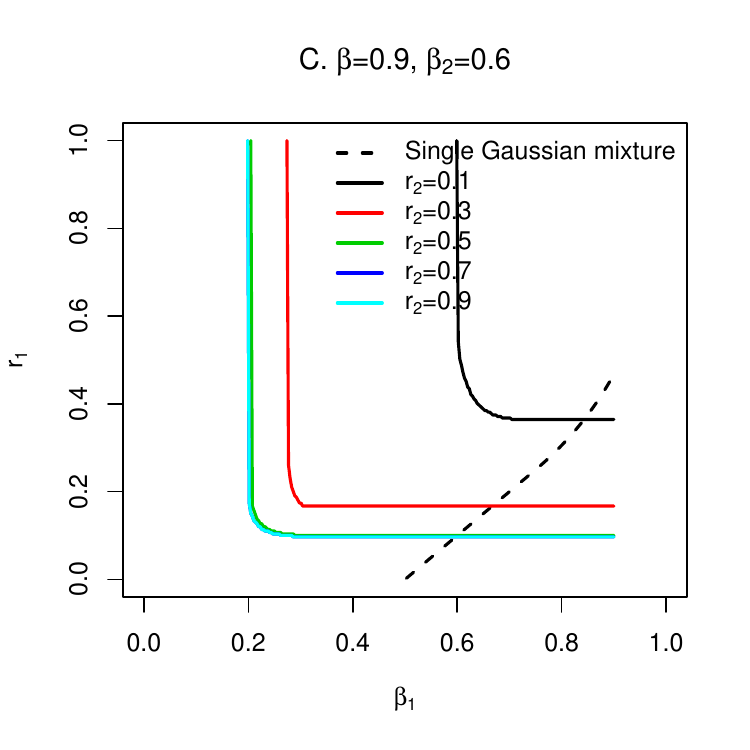}
  \caption{\label{fig:boundary}Detection boundary for normally distributed $T_{kj}$ following~\eqref{eq:normal}. Dotted line corresponds to the detection boundary for the single sequence of test statistics $T_{1j}$ when $\beta_1>1/2$. Panel A fixes $r_1=r_2$ and $\beta_1=\beta_2$. Panel B shows how the boundary varies with $\beta_2$, and panel C shows how it varies with $r_2$. The region below each colored line is the undetectable region.}
\end{figure}

For a concrete example of the detection boundary, suppose that
\begin{equation}
  \label{eq:normal}
  F^0_k\sim \mathcal{N}(0,1),
  \quad
  F^1_k\sim \mathcal{N}\{(2r_k\ln p)^{1/2},1\}
\end{equation}
for some positive constants $r_k$, $k=1,2$, which satisfies Assumptions~\ref{a:sto_ord} and \ref{a:tails}. The corresponding $\alpha$ functions, which appear in the inequalities from Theorem~\ref{thm:undetectable}, were presented above in~\eqref{eq:alpha_normal}. Then the detection boundary can be illustrated by plotting the boundary of the undetectable region. This is shown in Figure~\ref{fig:boundary} for various values of $\beta$, $\beta_k$, and $r_k$. It is interesting to compare these results to the boundary for detecting sparse mixtures in a single sequence of tests, e.g., testing $H_0:\pi_1=0$, which was computed under~\eqref{eq:normal} by \citet{donoho2004higher} and is plotted in Figure~\ref{fig:boundary}.

\subsection{\label{sec:implications}Implications}
Theorems~\ref{thm:adaptive}--\ref{thm:boundary} and Figure~\ref{fig:boundary} reveal a number of interesting features that decide the difficulty of testing weak positive latent dependence~\eqref{eq:test}. Most obviously, detection is easier for smaller $\beta$, corresponding to stronger dependence. It is also in general easier for larger $\alpha_k(a)$ and $v_k(x)$, which correspond to larger differences between the null and alternative distributions. To illustrate this, for normally distributed signals~\eqref{eq:normal} it can be shown that $v^+_k(x)=-(x^{1/2}-r_k^{1/2})_+^2$. For large signal strengths $r_k\geq1$, $v^+_k(x)=0$ on $x\in(0,1)$, so by Theorem~\ref{thm:boundary} and inequality \eqref{eq:Q1} the detectable region is
\[
0<
\sup_{\substack{x_1,x_2>0,\\x_1+x_2<1}}\left(\frac{1}{2}-\beta+\sum_{k=1}^2\frac{x_k\wedge\beta_k}{2}\right)
=
\frac{1}{2}-\beta+\frac{1\wedge(\beta_1+\beta_2)}{2}.
\]
This implies that for strong signals, and when the individual latent indicator sequences $I_{kj}$ are sufficiently sparse such that $\beta_1+\beta_2>1$, any $\epsilon=\pi_1\pi_2+p^{-\beta}$ is detectable. In this setting it would be more interesting to calibrate $\epsilon$ to approach $\pi_1\pi_2$ at faster than a polynomial rate.

Another implication is that for fixed $\beta$ and $\alpha_k(a)$, dependency detection is more difficult for smaller $\beta_k$. Even when $r_k\geq1$, the previous inequality shows that dependence may be undetectable if $\beta>(1+\beta_1+\beta_2)/2$. When there are many non-null signals in the two sequences of test statistics, many features with both $I_{1j}=1$ and $I_{2j}=1$ are necessary to provide significant evidence for dependence, even if the $I_{kj}$ were directly observed.

Finally, Figure~\ref{fig:boundary} reveals an interesting connection to the single-sequence sparse mixture detection problem. First, since signals must exist in both sequences of test statistics for there to exist dependence, a test for weak dependence such as~\eqref{eq:adaptive_test} can also be used as a method to detect sparse mixtures in a single sequence of test statistics. Second, panels B and C of Figure~\ref{fig:boundary} show that a portion of the undetectable region of the single-sequence problem lies within the detectable region of dependency detection. This means that the proposed test~\eqref{eq:adaptive_test} using $T_{1j}$ and $T_{2j}$ can actually detect signal in one of the sequences even when detection is theoretically impossible using that sequence alone. Intuitively, this can occur when the non-null signals of one sequence, say the $T_{2j}$, are strong enough to be easily identified. Then dependency could be detected simply by checking only the $T_{1j}$ paired with the non-null $T_{2j}$ to see if they are also non-null. This greatly reduces the dimensionality of the problem, and so could succeed even if the non-null signals in the $T_{1j}$ are so weak that they cannot be detected by single-sequence methods.

\section{\label{sec:sims}Numerical results}

\subsection{\label{sec:methods}Methods studied}
The proposed statistic $\widehat{\mathcal{D}}$~\eqref{eq:Dhat} was compared to several other existing procedures for testing~\eqref{eq:test}. Spearman's correlation is the most straightforward naive approach. Brownian distance covariance \citep{szekely2009brownian} is a recently developed nonparametric method designed for omnibus power. The GPA method \citep{chung2014gpa} was specifically developed for test statistics following model~\eqref{eq:mixture}, though it was designed for strong rather than weak dependence and makes parametric assumptions on the $T_{kj}$, namely $F^0_k\sim \mathcal{U}(0,1)$ and $F^1_k\sim \mathcal{B}(\alpha_1,\alpha_2)$. The $M^{DDP}_{m\times m}$ test of \citet{heller2016consistent} generalize several classical tests for independence. It calculates the Pearson chi-square test statistics for independence across all possible $m\times m$ contingency tables induced by the observed $(T_{1j},T_{2j})$, as illustrated in Figure~\ref{fig:2x2}, and aggregates them by taking their maximum. It then combines this max statistic across all $m=2,\ldots,M$. For computational reasons, in these simulations $M$ was set to equal 3. Finally, the method of \citet{zhao2017sparse}, referred to here as the max test, tests~\eqref{eq:test} using $\max_j\{\min(T_{1j},T_{2j})\}$ and provides a closed-formed expression for the permutation $p$-value. Two hundred permutations were used to calculate $p$-values for the $\widehat{\mathcal{D}}$, Brownian distance covariance, and $M^{DDP}_{m\times m}$ tests.

All simulations were conducted under a ``fixed-effect'' sampling scheme, where the non-null indicators $I_{kj}$ were generated once and then fixed across replications. This was done because in many applications, for example in statistical genomics, whether or not a genomic feature exhibits a non-null effect does not change across repeated sampling. To generate the $I_{kj}$, under $H_0$, $p\pi_k$ of the $I_{kj}$ were randomly set to 1, independently for $k=1$ and $k=2$. Under $H_A$, $p\epsilon$ of the features were randomly chosen to be simultaneously non-null in both sequences, with $I_{1j}=I_{2j}=1$, while maintaining a total of $p\pi_k$ non-null signals in each sequence. Finally, conditional on the $I_{kj}$, $T_{kj}$ were generated according to the mixture model~\eqref{eq:mixture}. All simulations were conducted under the rare and weak signal setting, as described in Section~\ref{sec:intro}, where the number of non-null signals, as well as their effect sizes, are small.

\subsection{\label{sec:ind}Independent tests}
These simulations consider test statistics $T_{kj}$ that are independent across $j$. Null and non-null signals were generated according to $T_{kj}\sim\vert \mathcal{N}(0,1)\vert$ and $T_{kj}\sim\vert \mathcal{N}(\mu_{kj},\sigma_{kj}^2)\vert$, respectively. To set the parameters of the non-null distribution, the $\mu_{kj}$ were generated from $\mathcal{N}(2.5,1)$ and the $\sigma_{kj}^2$ were generated from a Gamma distribution with shape equal to 2 and scale equal to 1. These parameters, like the $I_{kj}$, were generated once and then fixed across all replications. The total number of features, $p=10^3$, was relatively small in order to accommodate the computationally intensive nature of the distance covariance and $M^{DDP}_{m\times m}$ methods. The proposed statistic $\widehat{\mathcal{D}}$~\eqref{eq:Dhat} could therefore be calculated without using the truncated version described in Section~\ref{sec:implementation}. To implement GPA, which requires $p$-values as input, the $T_{kj}$ were transformed according to $2\Phi(-T_{kj})$.

\begin{table}
  \begin{center}
  \caption{\label{tab:ind_typeI}Empirical type I errors for $p=10^3$ independent tests at nominal significance level $\alpha=0.05$ over 400 replications. dcov = Brownian distance covariance; $M^{DDP}_{m\times m}$ = max aggregation method of \citet{heller2016consistent}; GPA = method of \citet{chung2014gpa}; Max = method of \citet{zhao2017sparse}; $\widehat{\mathcal{D}}$ = proposed method.}
  \begin{tabular}{rcccccc}
    & \multicolumn{6}{c}{Number of signals in sequences 1 and 2}\\
    & (5,5) &  (10,5) & (15,5) &  (10,10) & (15,10) & (15,15) \\
    \hline
    Spearman & 0.04 & 0.04 & 0.06 & 0.05 & 0.06 & 0.06 \\ 
    dcov & 0.05 & 0.05 & 0.07 & 0.05 & 0.06 & 0.05 \\ 
    $M^{DDP}_{m\times m}$ & 0.06 & 0.05 & 0.09 & 0.05 & 0.09 & 0.07 \\ 
    GPA & 0.02 & 0.01 & 0.01 & 0.01 & 0.01 & 0.01 \\ 
    Max & 0.07 & 0.04 & 0.02 & 0.04 & 0.03 & 0.03 \\ 
    $\widehat{\mathcal{D}}$ & 0.04 & 0.03 & 0.04 & 0.03 & 0.02 & 0.02 \\
    \hline
  \end{tabular}
  \end{center}
\end{table}

\begin{figure}
  \begin{center}
    \includegraphics[scale=0.5]{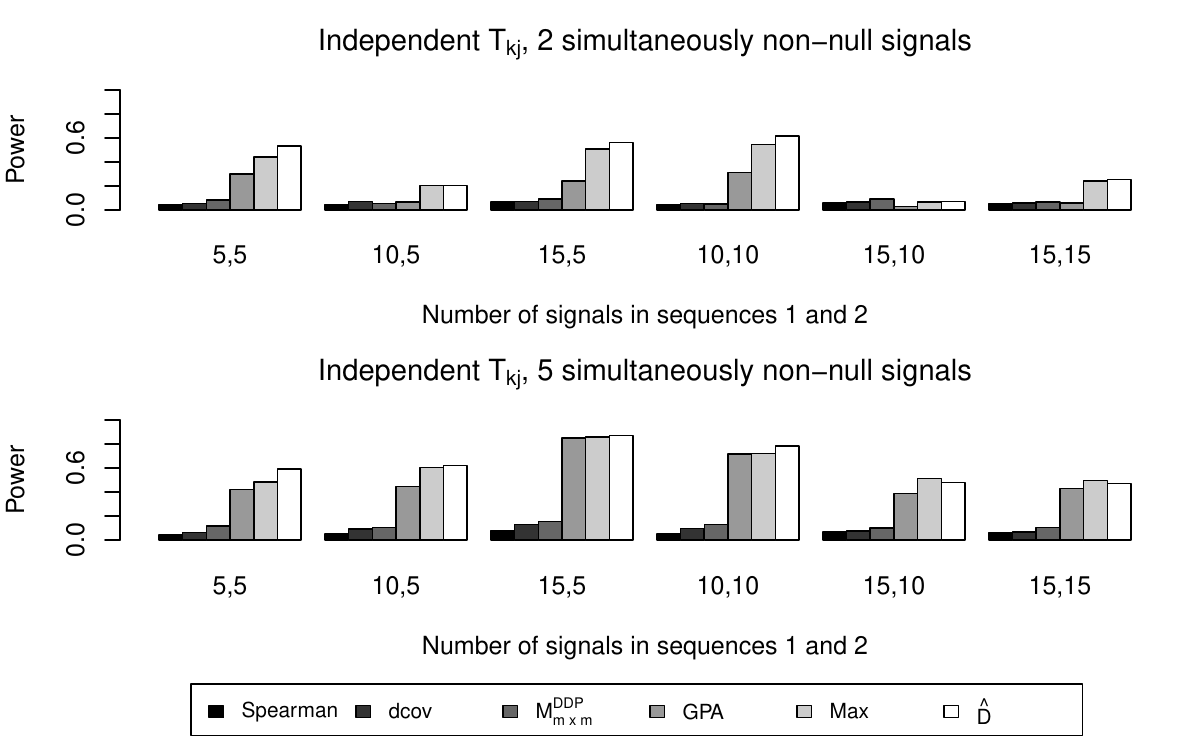}
  \caption{\label{fig:ind}Empirical powers for $p=10^3$ independent tests at nominal significance level $\alpha=0.05$ over 400 replications. dcov = Brownian distance covariance;  $M^{DDP}_{m\times m}$ = max aggregation method of \citet{heller2016consistent}; GPA = method of \citet{chung2014gpa}; Max = method of \citet{zhao2017sparse}; $\widehat{\mathcal{D}}$ = proposed method.}
  \end{center}
\end{figure}

Table~\ref{tab:ind_typeI} reports the empirical type I errors for simulation settings with different numbers of non-null tests in each sequence of test statistics.  The proposed method was able to control the type I error at the nominal $\alpha=0.05$ level. Figure~\ref{fig:ind} reports the empirical powers under various simulation settings. Detecting dependence was easier for all methods when there were more simultaneous signals, corresponding to smaller $\beta$ from calibration~\eqref{eq:calibrate}. The proposed $\widehat{\mathcal{D}}$ had the highest power in almost all settings. Figure~\ref{fig:ind_power_curve} in the Appendix plots the powers versus the number of simultaneous signals when there were 15 non-null signals in each sequence. GPA had the highest power under strong dependence, when there were many simultaneous signals, but $\widehat{\mathcal{D}}$ was the most powerful method under weak dependence. The proposed method was closely matched by the max test of \citet{zhao2017sparse} under weak dependence but outperformed the max test when there were more than 10 simultaneous signals.

\subsection{\label{sec:dep}Dependent tests}
These simulations generate $T_{kj}$ that are dependent across $j$. The total number of features was again $p=10^3$. Realistic correlation structures were generated using real genotype data from a randomly chosen set of $p$ adjacent genetic variants on human chromosome 1, obtained from the pediatric autoimmune disease data discussed in Section~\ref{sec:application}.

In each replication, $n=200$ subjects from these data were selected at random to serve as data from hypothetical study $k=1$, and another $n=200$ were independently selected to serve as data from hypothetical study $k=2$. To generate test statistics $T_{kj}$ from these studies, simulated outcomes $Y_k$ were first generated according to linear models $Y_k=S_k\theta_k+\varepsilon_k$, where the $S_k$ were $n\times p$ matrices of additively coded genotypes of all variants, the $\theta_k=(\theta_{k1},\ldots,\theta_{kp})^\top$ were $p\times1$ coefficient vectors, and the $\varepsilon_k$ were $n\times1$ vectors of independent standard normal errors. The $\theta_{kj}$ corresponding to variants with $I_{kj}=0$ were set to zero. The remaining non-zero $\theta_{kj}$, corresponding to variants with $I_{kj}=1$, were generated from $\mathcal{N}(0.5,0.2)$ and then randomly multiplied by either 1 or $-1$. All $\theta_{kj}$ were generated once and then fixed across all replications. Finally, the $T_{kj}$ were taken to be the absolute values of the $Z$-statistics for the marginal regressions of $Y_k$ on the $j$th variant.

\begin{table}
  \begin{center}
  \caption{\label{tab:dep_typeI}Empirical type I errors for $p=10^3$ dependent tests at nominal significance level $\alpha=0.05$ over 400 replications. dcov = Brownian distance covariance; $M^{DDP}_{m\times m}$ = max aggregation method of \citet{heller2016consistent}; GPA = method of \citet{chung2014gpa}; Max = method of \citet{zhao2017sparse}; $\widehat{\mathcal{D}}$ = proposed method.}
  \begin{tabular}{rcccccc}
    & \multicolumn{6}{c}{Number of signals in sequences 1 and 2}\\
    & (5,5) &  (10,5) & (15,5) &  (10,10) & (15,10) & (15,15) \\
    \hline
    Spearman & 0.23 & 0.26 & 0.21 & 0.24 & 0.22 & 0.26 \\ 
    dcov & 0.35 & 0.41 & 0.46 & 0.45 & 0.41 & 0.49 \\ 
    $M^{DDP}_{m\times m}$ & 0.46 & 0.51 & 0.60 & 0.56 & 0.54 & 0.64 \\ 
    GPA & 0.06 & 0.21 & 0.24 & 0.33 & 0.25 & 0.34 \\ 
    Max & 0.01 & 0.01 & 0.00 & 0.00 & 0.01 & 0.00 \\ 
    $\widehat{\mathcal{D}}$ & 0.04 & 0.05 & 0.03 & 0.01 & 0.03 & 0.04 \\
    \hline
  \end{tabular}
  \end{center}
\end{table}

\begin{figure}
  \begin{center}
    \includegraphics[scale=0.5]{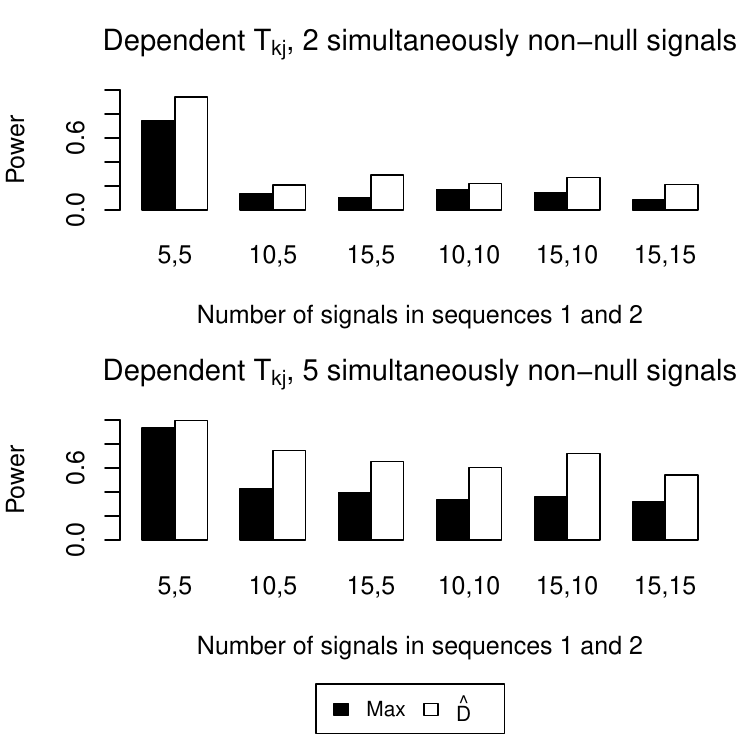}
  \caption{\label{fig:dep}Empirical powers for $p=10^3$ dependent tests at nominal significance level $\alpha=0.05$ over 400 replications. Max = method of \citet{zhao2017sparse}; $\widehat{\mathcal{D}}$ = proposed method.}
  \end{center}
\end{figure}

Table~\ref{tab:dep_typeI} reports the empirical type I errors under different simulation settings for dependent test statistics. It is interesting that the proposed $\widehat{\mathcal{D}}$, which uses the simple permutation procedure described in Section~\ref{sec:inf}, was still able to control the type I error in this setting. The only other method able to achieve this was the max test of \citet{zhao2017sparse}. Figure~\ref{fig:dep} reports the empirical powers and power curves of only those methods with proper type I error control. The proposed $\widehat{\mathcal{D}}$ was consistently more powerful than the max test. Figure~\ref{fig:dep_power_curve} in the Appendix plots the power curves as a function of the number of simultaneous signals, and $\widehat{\mathcal{D}}$ was the most powerful at all levels of dependence.

\subsection{\label{sec:alt}Alternative dependency detection procedures}
Several  variants of the compared dependency detection procedures were also explored. First, truncated versions of the proposed $\widehat{\mathcal{D}}$, described in Section~\ref{sec:implementation}, can be calculated with different truncation parameters $m_1$ and $m_2$. Next, instead of taking the maximum of the Pearson test statistics from all induced $m\times m$ tables, \citet{heller2016consistent} also proposed the sum aggregation test $S^{DDP}_{m\times m}$, which adds them. Finally, define the test statistic
\begin{equation}
  \label{eq:Dtilde}
  \widetilde{\mathcal{D}}
  =
  \sup_{(t_1,t_2)\in\mathcal{S}}p^{1/2}\frac{\vert\hat{S}_{12}(t_1,t_2)-\hat{S}_1(t_1)\hat{S}_2(t_2)\vert}{[\hat{S}_1(t_1)\{1-\hat{S}_1(t_2)\}\hat{S}_2(t_2)\{1-\hat{S}_2(t_2)\}]^{1/2}}.
\end{equation}
Unlike the denominator $\widehat{\mathcal{D}}$, which as discussed in Section~\ref{sec:teststat} favors tuples $(t_1,t_2)$ where both $t_1$ and $t_2$ are large, the denominator of $\widetilde{\mathcal{D}}$ gives higher weights whenever both $t_1$ and $t_2$ are both extreme, regardless of whether they are extremely large or extremely small. This denominator also makes $\widetilde{\mathcal{D}}$ closely related to the maximum of the square roots of Pearson chi-square test statistics \citep{thas2004nonparametric}. Two hundred permutations were used to calculate $p$-values for each of these methods.

\begin{table}
  \begin{center}
  \caption{\label{tab:ind_detailed_typeI}Empirical type I errors for $p=10^3$ independent tests at nominal significance level $\alpha=0.05$ over 400 replications for variations of the procedures. $M^{DDP}_{m\times m}$ = max aggregation method of \citet{heller2016consistent}; $S^{DDP}_{m\times m}$ = sum aggregation method of \citet{heller2016consistent}; $\widetilde{\mathcal{D}}$ = statistic~\eqref{eq:Dtilde}; $\widehat{\mathcal{D}}_{x}$ = truncated version of the proposed method with $m_1=m_2=x$; $\widehat{\mathcal{D}}$ = proposed method without truncation.}
  \begin{tabular}{rcccccc}
    & \multicolumn{6}{c}{Number of signals in sequences 1 and 2}\\
    & (5,5) &  (10,5) & (15,5) &  (10,10) & (15,10) & (15,15) \\
    \hline
    $M^{DDP}_{m\times m}$ & 0.06 & 0.05 & 0.09 & 0.05 & 0.09 & 0.07 \\ 
    $S^{DDP}_{m\times m}$ & 0.05 & 0.04 & 0.05 & 0.04 & 0.05 & 0.04 \\ 
    $\widetilde{\mathcal{D}}$ & 0.03 & 0.05 & 0.04 & 0.04 & 0.02 & 0.05 \\ 
    $\widehat{\mathcal{D}}_{10}$ & 0.04 & 0.05 & 0.03 & 0.03 & 0.01 & 0.03 \\ 
    $\widehat{\mathcal{D}}_{100}$ & 0.04 & 0.03 & 0.04 & 0.03 & 0.02 & 0.02 \\ 
    $\widehat{\mathcal{D}}$ & 0.04 & 0.03 & 0.04 & 0.03 & 0.02 & 0.02 \\
    \hline
  \end{tabular}
  \end{center}
\end{table}

\begin{figure}
  \begin{center}
    \includegraphics[scale=0.5]{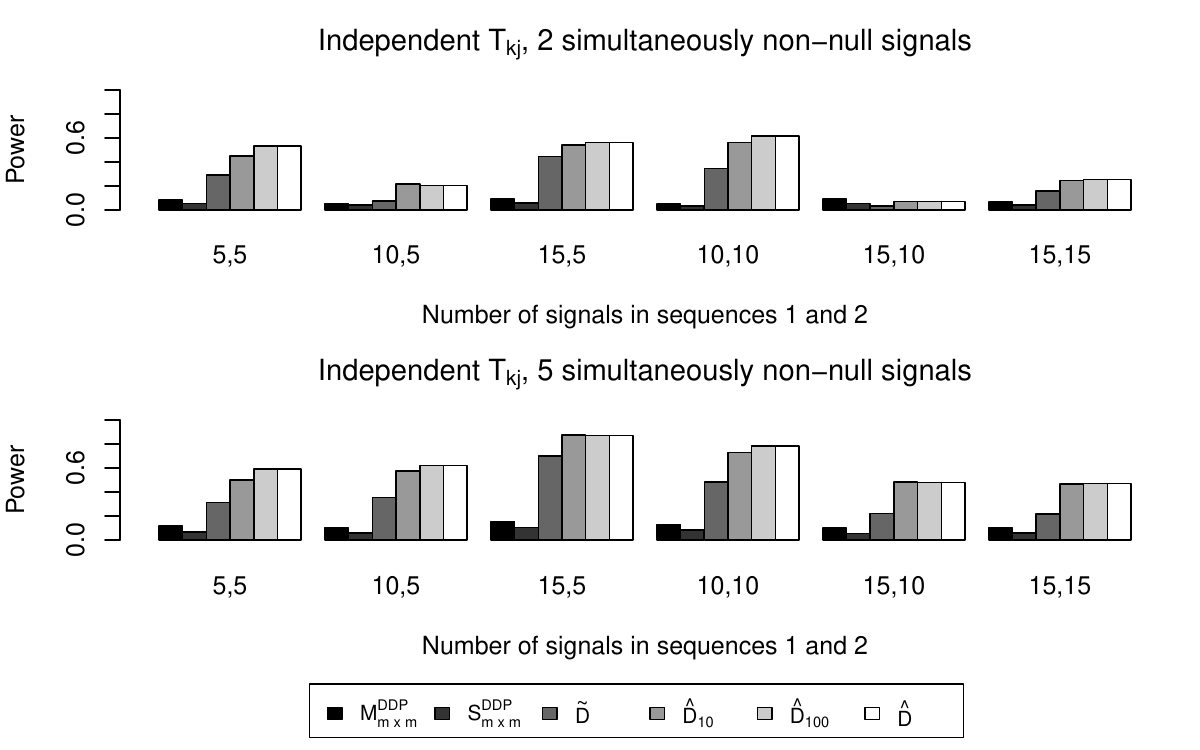}
  \caption{\label{fig:ind_detailed}Empirical powers for $p=10^3$ independent tests at nominal significance level $\alpha=0.05$ over 400 replications for variations of the procedures. $M^{DDP}_{m\times m}$ = max aggregation method of \citet{heller2016consistent}; $S^{DDP}_{m\times m}$ = sum aggregation method of \citet{heller2016consistent}; $\widetilde{\mathcal{D}}$ = statistic~\eqref{eq:Dtilde}; $\widehat{\mathcal{D}}_{x}$ = truncated version of the proposed method with $m_1=m_2=x$; $\widehat{\mathcal{D}}$ = proposed method without truncation.}
  \end{center}
\end{figure}

These variations were applied to the independent test statistic simulations from Section~\ref{sec:ind}. Table~\ref{tab:ind_detailed_typeI} indicates that most were able to maintain the nominal type I error rate. From Figure~\ref{fig:ind_detailed}, and the power curves in Figure~\ref{fig:ind_detailed_power_curve} in the Appendix, $\widehat{\mathcal{D}}$ was the best performer and had the same power as the truncated $\widehat{\mathcal{D}}_{100}$, which had truncation parameters $m_1=m_2=100$. The more heavily truncated $\widehat{\mathcal{D}}_{10}$, with $m_1=m_2=10$, was slightly less powerful, especially in the presence of a large number of simultaneous signals, but was still the best of the remaining procedures. This suggests that significant computational speedup can be achieved without sacrificing much power. The modified statistic $\widetilde{\mathcal{D}}$ was the next best performer. Max aggregation $M^{DDP}_{m\times m}$ always outperformed sum aggregation $S^{DDP}_{m\times m}$ and had more power than $\widetilde{\mathcal{D}}$ under strong dependence.

\subsection{\label{sec:single}Detection of single-sequence sparse mixture}
As discussed in Section~\ref{sec:implications} and illustrated in Figure~\ref{fig:boundary}, one implication of the detection boundary results is that dependency detection is sometimes possible when single-sequence signal detection is not. This section studies this phenomenon in simulations using $p=10^5$ pairs of test statistics.

The number of non-null signals in the first sequence of test statistics was either 282, 100 or 32. This corresponds to $\beta_1$ from~\eqref{eq:calibrate} equal to either 0.51, 0.6, or 0.7. The $T_{1j}$ were generated following
\[
T_{1j}\mid I_{1j}=0\sim \vert \mathcal{N}(0,1)\vert,\quad T_{1j}\mid I_{1j}=1\sim\vert \mathcal{N}[\{(2\beta_1-1)\ln p\}^{1/2},1]\vert.
\]
Existing results for the single-sequence detection problem imply that for these $T_{1j}$, it is impossible to detect the presence of non-null signals using single-sequence detection methods \citep{cai2011optimal,cai2014optimal,donoho2004higher,ingster1997some}. The second sequence of test statistics was generated with 316 non-null signals, corresponding to $\beta_2=0.5$. The $T_{2j}$ followed
\[
T_{2j}\mid I_{2j}=0\sim \vert \mathcal{N}(0,1)\vert,\quad T_{2j}\mid I_{2j}=1\sim\vert \mathcal{N}\{(2\ln p)^{1/2},1\}\vert,
\]
so that the non-null signals were very strong. Finally, under $H_A$, the dependency parameter $\beta$ was set to $\beta_1\vee\beta_2+0.01$, corresponding to either 251, 89, or 28 signals that were non-null in both sequences.

The distance covariance method of \citet{szekely2009brownian} and the $M^{DDP}_{m\times m}$ test of \citet{heller2016consistent} were not implemented for computational reasons, as $p$ is quite large in these simulations. The remaining dependency detection procedures were applied for the purpose of testing $H_0:\pi_1=0$. For comparison, the higher criticism method was also applied to the $T_{1j}$. \citet{donoho2004higher} showed that for these simulation settings, the higher criticism statistic is asymptotically adaptively optimal among all single-sequence detection methods. Its null distribution was approximated using 200 simulated realizations of $p$ standard normals, and this distribution was used to provide $p$-values.

\begin{table}
  \begin{center}
  \caption{\label{tab:III}Empirical type I errors and powers for single-sequence signal detection for $p=10^5$ tests at nominal significance level $\alpha=0.05$ over 400 replications. HC = higher criticism method of \citet{donoho2004higher}; GPA = method of \citet{chung2014gpa}; Max = method of \citet{zhao2017sparse}; $\widehat{\mathcal{D}}$ = proposed method.}
  \begin{tabular}{r|ccc|ccc}
    & \multicolumn{3}{c}{Type I errors} & \multicolumn{3}{c}{Powers}\\
    $\beta_1$: & 0.51 & 0.6 & 0.7 & 0.51 & 0.6 & 0.7\\ 
    \hline
    HC & --- &--- &--- & 0.04 & 0.07 & 0.14 \\ 
    Spearman & 0.04 & 0.04 & 0.06 & 0.04 & 0.07 & 0.07 \\ 
    GPA & 0.00 & 0.00 & 0.00 & 0.00 & 0.46 & 0.20 \\ 
    Max & 0.06 & 0.05 & 0.05&  0.13 & 0.63 & 0.80  \\ 
    $\widehat{\mathcal{D}}$ & 0.06 & 0.06 & 0.05 & 0.14 & 0.61 & 0.72 \\ 
    \hline
  \end{tabular}
  \end{center}
\end{table}

Table~\ref{tab:III} reports the empirical type I errors and powers for different values of $\beta_1$. The type I error refers to the null hypothesis of independence between $T_{1j}$ and $T_{2j}$, so it does not apply to higher criticism because it does not test independence. The proposed $\widehat{\mathcal{D}}$ and the max test of \citet{zhao2017sparse} both had substantial power to detect dependence, and thus to detect signal in $T_{1j}$, even when higher criticism did not.

\subsection{\label{sec:application}Application to pediatric autoimmune disease}
Different autoimmune diseases can be genetically related, meaning that there are genetic variants which are associated with more than one disease. The proposed $\widehat{\mathcal{D}}$ can be used to rigorously test the degree to which a pair of conditions are genetically related. Let $P_{kj}$ be the $p$-value for association between the $j$th variant and the $k$th disease, $k=1,2$. Testing for weak positive latent dependence~\eqref{eq:test} between the $P_{kj}$ is equivalent to testing whether there are more markers that affect both diseases than expected by chance.

Hakonsarson and colleagues at the Children's Hospital of Pennsylvania conducted separate genome-wide association studies in 10,718 shared controls and over 5,000 cases across ten different diseases: ankylosing spondylitis, Celiac's disease, common variable immunodeficiency, Crohn's disease, juvenile idiopathic arthritis, psoriasis, systemic lupus erythematosus, thyroiditis, type I diabetes, and ulcerative colitis \citep{li2015meta,li2015genetic}. Subjects were genotyped on Illumina Infinium HumanHap550 and Human610 BeadChip array platforms, and only variants common to both arrays and surviving quality control were used for analysis.

Only autosomal chromosomes were considered, and variants in the major histocompatibility complex region, defined as the 25,500,000 to 34,000,000 base pair region of chromosome 6, were not considered because they are known to be highly associated with all autoimmune diseases. This resulted in roughly 450,000 typed variants for each disorder. Genome-wide association $p$-values $P_{kj}$ were calculated for each variant. The correlation between test statistics from different studies due to the shared controls was found, using the method of \citet{zaykin2010p}, to be at most only 0.019.

The proposed $\widehat{\mathcal{D}}$~\eqref{eq:Dhat} was implemented with $m_1=m_2=1000$, with the truncation parameters $m_k$ defined in Section~\ref{sec:implementation}. The results were compared to those of Spearman's correlation, the GPA method of \citet{chung2014gpa}, and the max test of \citet{zhao2017sparse}. For computational reasons, the distance covariance method of \citet{szekely2009brownian} and the $M^{DDP}_{m\times m}$ test of \citet{heller2016consistent} were omitted. For $\widehat{\mathcal{D}}$ and the max test, the $P_{kj}$ were converted to $-\log_{10}P_{kj}$ in order to satisfy the stochastic ordering condition of Assumption~\ref{a:sto_ord}.

These methods were applied to test for weak positive latent dependence~\eqref{eq:test} between all 45 unique pairs of the 10 disorders. Permutation $p$-values for $\widehat{\mathcal{D}}$ were calculated using 10,000 random permutations; this procedure still maintain type I error in the presence of linkage disequilibrium, as shown in simulations in Section~\ref{sec:dep}. A Bonferroni correction can be applied to adjust for multiple comparisons, but this may be overly conservative because the pairwise nature of the 45 tests makes them highly dependent. As an alternative, it has been found that in this pairwise testing setting, the Benjamini-Hochberg procedure \citep{benjamini1995controlling} can still maintain false discovery rate control in practice \citep{yekutieli2008false}.

\begin{table}
  \begin{center}
    \caption{\label{tab:paid}Pairs of pediatric autoimmune diseases for which at least one testing method was significant at the 0.05 level after Bonferroni or Benjamini-Hochberg (BH) correction. Bold $p$-values are less than $0.05/45$, and bold BH-corrected $p$-values are less than $0.05$. Disorders: AS = ankylosing spondylitis; CEL = Celiac's disease; CD = Crohn's disease; CVID = common variable immunodeficiency; JIA = juvenile idiopathic arthritis; SLE = systematic lupus erythematosus; T1D = type I diabetes; THY = thyroiditis; UC = ulcerative colitis. Methods: GPA = method of \citet{chung2014gpa}; Max = method of \citet{zhao2017sparse}; $\widehat{\mathcal{D}}$ = proposed method.}
    \begin{tabular}{l|cccc|cccc}
      & \multicolumn{4}{c}{$p$-values} & \multicolumn{4}{c}{BH-corrected $p$-values} \\
      Disorders & Spearman & GPA & Max & $\widehat{\mathcal{D}}$ & Spearman & GPA & Max & $\widehat{\mathcal{D}}$\\
      \hline
      UC-CD & {\bf0.0000} & {\bf0.0000} & {\bf0.0000} & {\bf0.0001} & {\bf0.0000} & {\bf0.0000} & {\bf0.0004} & {\bf0.0015} \\ 
      CVID-JIA & 0.0137 & {\bf0.0000} & {\bf0.0000} & {\bf0.0001} & 0.0883 & {\bf0.0000} & {\bf0.0002} & {\bf0.0015} \\ 
      UC-JIA & 0.8101 & {\bf0.0004} & {\bf0.0000} & {\bf0.0001} & 0.8867 & {\bf0.0033} & {\bf0.0002} & {\bf0.0015} \\ 
      UC-T1D & {\bf0.0007} & {\bf0.0000} & {\bf0.0002} & {\bf0.0002} & {\bf0.0080} & {\bf0.0000} & {\bf0.0022} & {\bf0.0022} \\ 
      T1D-JIA & 0.0519 & 0.0060 & {\bf0.0006} & {\bf0.0003} & 0.1826 & {\bf0.0386} & {\bf0.0042} & {\bf0.0027} \\ 
      JIA-CD & 0.0568 & 0.0014 & {\bf0.0001} & {\bf0.0004} & 0.1826 & {\bf0.0102} & {\bf0.0008} & {\bf0.0030} \\ 
      T1D-CD & {\bf0.0000} & {\bf0.0000} & 0.0033 & 0.0013 & {\bf0.0000} & {\bf0.0000} & {\bf0.0211} & {\bf0.0079} \\ 
      THY-T1D & 0.6735 & 0.9232 & 0.0618 & 0.0014 & 0.8022 & 1.0000 & 0.1987 & {\bf0.0079} \\ 
      AS-CVID & 0.1494 & 1.0000 & 0.2023 & 0.0046 & 0.3169 & 1.0000 & 0.4791 & {\bf0.0225} \\ 
      AS-JIA & 0.0477 & 0.9631 & 0.2369 & 0.0050 & 0.1826 & 1.0000 & 0.5330 & {\bf0.0225} \\ 
      THY-SLE & 0.2059 & 0.9864 & 0.0155 & 0.0087 & 0.4028 & 1.0000 & 0.0871 & {\bf0.0356} \\ 
      THY-JIA & 0.0036 & 1.0000 & 0.6666 & 0.5730 & {\bf0.0273} & 1.0000 & 0.8228 & 0.8318 \\ 
      CVID-CD & {\bf0.0000} & 1.0000 & 0.9905 & 0.7264 & {\bf0.0006} & 1.0000 & 0.9905 & 0.8795 \\ 
      CEL-CD & {\bf0.0026} & 0.9005 & 0.9330 & 0.7965 & {\bf0.0230} & 1.0000 & 0.9905 & 0.8795 \\
      \hline
    \end{tabular}
  \end{center}
\end{table}

Table~\ref{tab:paid} presents disease pairs for which at least one dependency detection method was significant at an error rate of $0.05$ after either Bonferroni or Benjamini-Hochberg correction. The proposed $\widehat{\mathcal{D}}$ and the max test of \citet{zhao2017sparse} identified the most disease pairs after Bonferroni correction, while $\widehat{\mathcal{D}}$ alone gave the most findings after Benjamini-Hochberg correction. These results suggest that for detecting weak dependence, the test proposed in this paper is a valuable alternative to existing methods.

\begin{figure}
  \centering
  \includegraphics[height=0.85\textheight]{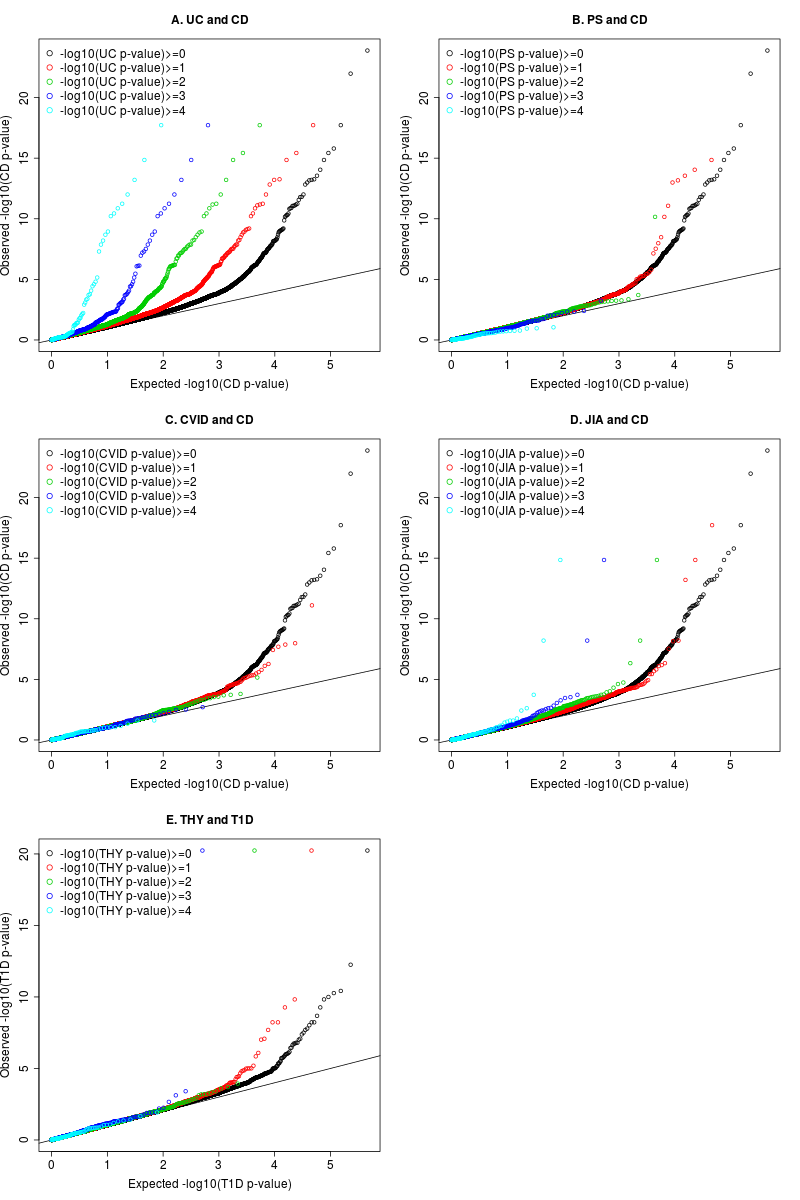}
  \caption{\label{fig:paid}Stratified QQ-plots of selected pairs of diseases. Diseases: CD = Crohn's disease; CVID = common variable immunodeficiency; PS = psoriasis; T1D = type I diabetes; THY = thyroiditis; UC = ulcerative colitis.}
\end{figure}

Figure~\ref{fig:paid} illustrates some selected results. For each disorder pair, it depicts different QQ-plots of the $-\log_{10}p$-values of one of the disorders, for those variants that have $-\log_{10}p$-values of at least certain sizes in the other disorder. Panel A illustrates the ulcerative colitis--Crohn's disease pair. As both are inflammatory bowel diseases, it is no surprise that these were found by all methods to exhibit genetic sharing. Panel A indeed shows that genetic variants that are more significant in the ulcerative colitis genome-wide association study also tend to be more significant in the Crohn's disease study.

In contrast, panel B illustrates the psoriasis--Crohn's disease pair, which was one of the pairs not found to be significant by any method. The QQ-plots reflect the fact that variants more significant in one study are not always more significant in the other. Panel C illustrates the common variable immunodeficiency--Crohn's disease pair, which was found to be significant only by Spearman's test. The QQ-plots show a negative dependence, which is not of interest here.

Finally, panel D illustrates the juvenile idiopathic arthritis--Crohn's disease pair, which was detected only by the max test of \citet{zhao2017sparse} and the proposed $\widehat{\mathcal{D}}$ after Bonferroni correction. Panel E illustrates the thyroiditis--type I diabetes pair, which was detected only by $\widehat{\mathcal{D}}$ after Benjamini-Hochberg correction. Both pairs of QQ-plots show the presence of positive dependence, but in contrast to the type of strong positive dependence present in panel A, the dependence here appears to be heavily driven by small number of variants that are significant in both disorders. This is exactly the type of weak dependence that is difficult for existing methods to detect, and exactly the motivation behind the method proposed in this paper.

\section{\label{sec:discussion}Discussion}
The simulations in Section~\ref{sec:sims} considered only the particular type of dependence described in equations~\eqref{eq:mixture} and~\eqref{eq:test}. The proposed method and the competing procedures have different properties otherwise. For example, when the $(T_{1j},T_{2j})$ are dependent in such a way so as to form a circle when plotted in $\mathbb{R}^2$, a small scale simulation study with 100 replications showed that both $M^{DDP}_{m\times m}$ and $S^{DDP}_{m\times m}$ had 100\% power, $\widetilde{\mathcal{D}}$~\eqref{eq:Dtilde} had 76\% power, and the proposed $\widehat{\mathcal{D}}$ had only 6\% power. Thus while $\widehat{\mathcal{D}}$ was the best performer under the dependence alternative considered in this paper, it is not an omnibus test for independence. Interestingly, its generalization $\widetilde{\mathcal{D}}$ had good all-around performance, making it a potentially good candidate for detecting general dependence alternatives.

The asymptotic properties of the proposed method were derived in Section~\ref{sec:theory} under the assumption that the test statistics $T_{kj}$ are independent across $j$. When the $T_{kj}$ are correlated, $\widehat{\mathcal{D}}$~\eqref{eq:Dhat} will likely no longer be asymptotically optimal, though the simulations in Section~\ref{sec:dep} indicate that it can still have good power. \citet{hall2008properties} studied the asymptotic properties of the higher criticism procedure for single-sequence signal detection with correlated tests, and \citet{hall2010innovated} proposed the innovated higher criticism method that can achieve optimality for certain correlation structures. However, their results do not immediately extend to testing for dependence~\eqref{eq:test}, and further work is necessary to determine the fundamental limits of detection as well as to develop optimal methods.
 
Some alternatives to the proposed $\widehat{\mathcal{D}}$ may have better finite-sample performance. Recently, \citet{li2015higher} showed that for the single-sequence detection problem, a test based on the Berk-Jones goodness-of-fit statistic can be dramatically more powerful than the higher criticism statistic, on which $\widehat{\mathcal{D}}$ is based. Previously, \citet{jager2007goodness} showed that the Berk-Jones-based test has the same asymptotic optimality properties as higher criticism. A similar statistic for testing~\eqref{eq:test} would be a useful alternative to $\widehat{\mathcal{D}}$.

This paper assumes that the $T_{kj}$ are two-tailed test statistics, and as such ignores the directions of effect of the non-null signals. However, it may be desirable to require variants to exhibit the same directions of effect in order to be considered as evidence for genetic sharing. There exist methods that can test for this more stringent condition \citep{heller2014replicability}, and it would be interesting to study their asymptotic properties.

\section*{Acknowledgments}
The authors thank Drs. Hakon Hakonarson, Brendan J. Keating, Yun Li, and Julie Kobie for providing the pediatric autoimmune disease data and helping with its analysis, Dr. Yihong Wu for helpful discussions, and the anonymous referees for excellent suggestions. The research of Tony Cai was supported in part by National Science Foundation grants DMS-1208982 and DMS-1403708, and the National Institutes of Health grant R01 CA127334. The research of Hongzhe Li was supported in part by the National Institutes of Health grants R01 GM097505 and R01 CA127334. The research of Dave Zhao was supported in part by National Science Foundation grant DMS-1613005 and the Simons Foundation grant SFLife 291812.

\bibliographystyle{abbrvnat}
\bibliography{refs}

\appendix
\section{\label{sec:appendix}Appendix}

\subsection{\label{sec:add_sim_res}Additional simulation results}
\begin{figure}
  \begin{center}
    \includegraphics[scale=0.5]{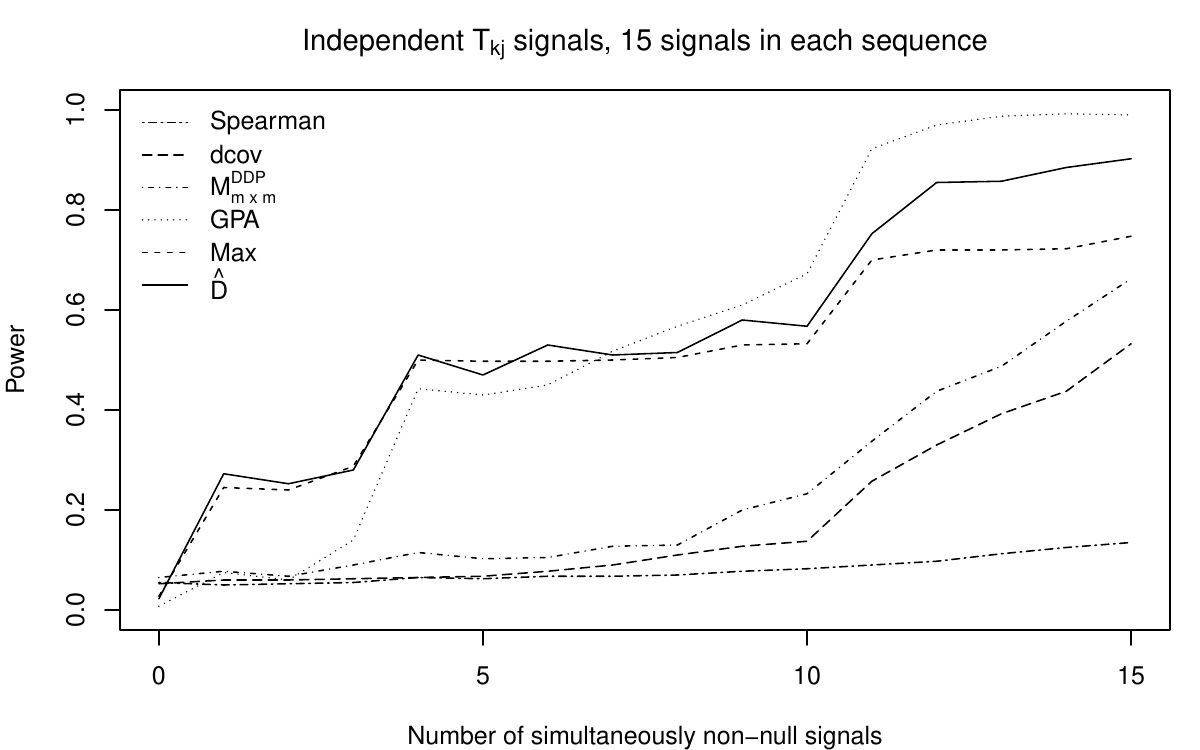}
  \caption{\label{fig:ind_power_curve}Power curves for $p=10^3$ independent tests at nominal significance level $\alpha=0.05$ over 400 replications. dcov = Brownian distance covariance;  $M^{DDP}_{m\times m}$ = max aggregation method of \citet{heller2016consistent}; GPA = method of \citet{chung2014gpa}; Max = method of \citet{zhao2017sparse}; $\widehat{\mathcal{D}}$ = proposed method.}
  \end{center}
\end{figure}

\begin{figure}
  \begin{center}
    \includegraphics[scale=0.5]{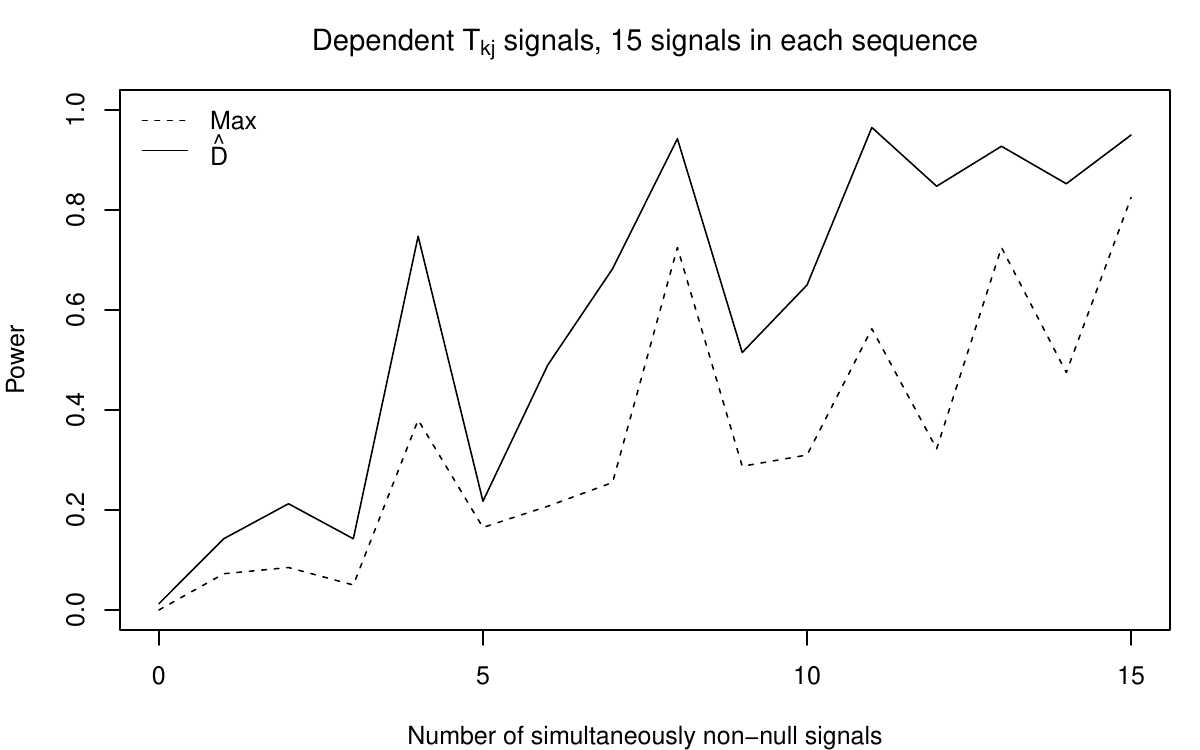}
  \caption{\label{fig:dep_power_curve}Power curves for $p=10^3$ dependent tests at nominal significance level $\alpha=0.05$ over 400 replications. Max = method of \citet{zhao2017sparse}; $\widehat{\mathcal{D}}$ = proposed method.}
  \end{center}
\end{figure}

\begin{figure}
  \begin{center}
    \includegraphics[scale=0.5]{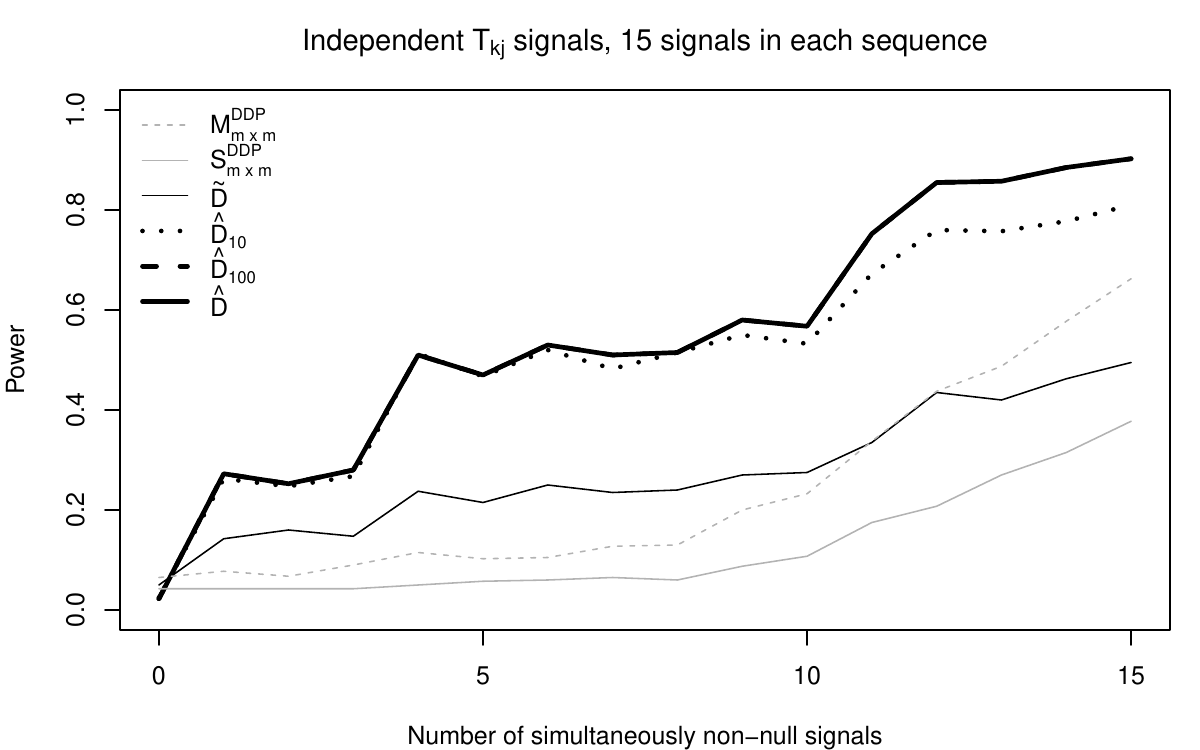}
  \caption{\label{fig:ind_detailed_power_curve}Power curves for $p=10^3$ independent tests at nominal significance level $\alpha=0.05$ over 400 replications for variations of the procedures. $M^{DDP}_{m\times m}$ = max aggregation method of \citet{heller2016consistent}; $S^{DDP}_{m\times m}$ = sum aggregation method of \citet{heller2016consistent}; $\widetilde{\mathcal{D}}$ = statistic~\eqref{eq:Dtilde}; $\widehat{\mathcal{D}}_{x}$ = truncated version of the proposed method with $m_1=m_2=x$; $\widehat{\mathcal{D}}$ = proposed method without truncation.}
  \end{center}
\end{figure}

Figures~\ref{fig:ind}--\ref{fig:ind_detailed} in the main text report the power of the proposed $\widehat{\mathcal{D}}$, competing methods, and several variations of these procedures for either 2 or 5 simultaneous signals. Below, Figures~\ref{fig:ind_power_curve}--\ref{fig:ind_detailed_power_curve} explore how the powers are affected by the number of simultaneous signals when there were 15 non-null signals in each sequence.

Figure~\ref{fig:ind_power_curve} considers the independent test statistics studied in Figure~\ref{fig:ind} of Section~\ref{sec:ind} in the main text. It shows that GPA was the best performer under strong dependence, when a large proportion of the 15 non-null signals were simultaneous signals. On the other hand, in the weak dependence regime of interest in this paper, $\widehat{\mathcal{D}}$ had the highest power of all methods. Its performance was similar to that of the max test of \citet{zhao2017sparse} under weak dependence but was superior under strong dependence.

Figure~\ref{fig:dep_power_curve} considers the dependent test statistics studied in Figure~\ref{fig:dep} of Section~\ref{sec:dep} in the main text. These power curves were not as smooth as those in Figure~\ref{fig:ind_power_curve} for the independent tests, in part because with dependent test statistics the power is a function of not just the number of simultaneous signals, but also specifically where those signals are located relative to the covariance structure. The proposed $\widehat{\mathcal{D}}$ dominated the max test of \citet{zhao2017sparse} in terms of power.

Figure~\ref{fig:ind_detailed_power_curve} considers variants of the dependency detection procedures applied to the independent test statistics, as in Figure~\ref{fig:ind_detailed} in Section~\ref{sec:alt} in the main text. As expected, the non-truncated $\widehat{\mathcal{D}}$ had the best performance. Somewhat surprisingly, the heavily truncated $\widehat{\mathcal{D}}_{10}$ had very good performance as well. The $M^{DDP}_{m\times m}$ test of \citet{heller2016consistent} outperformed the max test of \citet{zhao2017sparse} under strong dependence, but otherwise it and the $S^{DDP}_{m\times m}$ test of \citet{heller2016consistent} had the lowest powers.

\subsection{\label{sec:lemmas}Lemmas}
Some useful results from \citet{einmahl1985bounds} and \citet{cai2014optimal} are reproduced here for completeness.

\begin{lemma}[\citet{einmahl1985bounds} Corollary~2]
  \label{l:lil}
  Let $\mathcal{D}$ be defined as in \eqref{eq:oracle}. Under $H_0$ of~\eqref{eq:test},
  \[
  \limsup_{p\rightarrow\infty}\frac{\ln\mathcal{D}}{\ln\ln p}\stackrel{a.s.}{=}1.
  \]
\end{lemma}

\begin{lemma}[\citet{cai2014optimal} Lemma~3]
  \label{l:ess}
  Let $(X,\mathcal{F},\nu)$ be a measure space. Let $F:X\times\mathbb{R}_+\rightarrow\mathbb{R}_+$ be measurable. Assume that
  \[
  \lim_{M\rightarrow\infty}\frac{\ln F(x,M)}{M}=f(x)
  \]
  holds uniformly in $x\in X$ for some measurable $f:X\rightarrow\mathbb{R}$. If
  \[
  \int_X\exp(M_0f)d\nu<\infty
  \]
  for some $M_0>0$, then
  \[
  \lim_{M\rightarrow\infty}\frac{1}{M}\ln\int_XF(x,M)d\nu=\esssup_{x\in X}f(x).
  \]
\end{lemma}

The following lemmas are used to prove the results in this paper.

\begin{lemma}
  \label{l:cdf}
  Under Assumption~\ref{a:tails}, for $x\geq\log_p2$,
  \begin{align*}
    &F^1_k\{(F^0_k)^{-1}(p^{-x})\}=p^{v^-_k(x)+o(1)},\\
    &F^1_k\{(F^0_k)^{-1}(1-p^{-x})\}=1-p^{v^+_k(x)+o(1)},
  \end{align*}
  where
  \[
  v^-_k(x)=\esssup_{a\geq x}\{\alpha^-_k(a)-a\},
  \quad
  v^+_k(x)=\esssup_{a\geq x}\{\alpha^+_k(a)-a\}.
  \]
\end{lemma}
\textit{Proof}. When $x\geq\log_p2$, making the change of variables $u\mapsto(F^0_k)^{-1}(p^{-a})$ implies
\[
du
=
-\frac{F^0_k\{(F^0_k)^{-1}(p^{-a})\}\ln p}{f^0_k\{(F^0_k)^{-1}(p^{-a})\}}da
=
-\frac{p^{-a}\ln p}{f^0_k\{(F^0_k)^{-1}(p^{-a})\}}da.
\]
Therefore
\begin{align*}
  F^1_k\{(F^0_k)^{-1}(p^{-x})\}
  =\,&
  \int_{-\infty}^{(F^0_k)^{-1}(p^{-x})}f^1_k(u)du
  =
  -\ln p\int_{\infty}^x\exp[\ell_k\{(F^0_k)^{-1}(p^{-a})\}]p^{-a}da\\
  =\,&
  p^{o(1)}\int_x^\infty\exp[\ell_k\{(F^0_k)^{-1}(p^{-a})\}-a\ln p]da,
\end{align*}
where $\ell_k$ is the log-likelihood ratio $\ln(f_k^1/f_k^0)$. Defining
\[
F(x,M)=\exp[\ell_k\{(F^0_k)^{-1}(p^{-a})\}-aM]
\]
and $M=\ln p$, Assumption~\ref{a:tails} and Lemma~\ref{l:ess} imply that
\[
(\ln p)^{-1}\ln\int_x^\infty F(x,\ln p)da=\esssup_{x\leq a}\{\alpha^-_k(a)-a\}+o(1),
\]
Therefore
\[
F^1_k\{(F^0_k)^{-1}(p^{-x})\}
=
p^{\esssup_{x\leq a}\{\alpha_k^-(a)-a\}+o(1)}.
\]

Similarly, making the change of variables $u\mapsto (F^0_k)^{-1}(1-p^{-a})$ implies
\[
du
=
\frac{p^{-a}\ln p}{f^0_k\{(F^0_k)^{-1}(p^{-a})\}}da.
\]
Therefore
\begin{align*}
  F^1_k\{(F^0_k)^{-1}(1-p^{-x})\}
  =&\,
  \int_{-\infty}^{(F^0_k)^{-1}(1-p^{-x})}f^1_k(u)du
  =
  1-\int_{(F^0_k)^{-1}(1-p^{-x})}^\infty f^0_k(u)du\\
  =&\,
  1-\ln p\int_x^\infty\exp[\ell_k\{(F^0_k)^{-1}(1-p^{-a})\}]p^{-a}da\\
  =&\,
  1-p^{\esssup_{x\leq a}\{\alpha^+_k(a)-a\}+o(1)}.
\end{align*}

\begin{lemma}
  \label{l:v<0}
  Under Assumption~\ref{a:tails},
  \[
  \esssup_{a\geq\log_p2}\{\alpha^-_k(a)-a\}\leq0,
  \quad
  \esssup_{a\geq\log_p2}\{\alpha^+_k(a)-a\}\leq0.
  \]
\end{lemma}
\textit{Proof}. The changes of variables $u\mapsto(F^0_k)^{-1}(p^{-a})$ and $v\mapsto(F^0_k)^{-1}(1-p^{-a})$ imply that
\begin{align*}
  1
  =\,&
  \int_{-\infty}^{(F^0_k)^{-1}(0.5)}f^1_k(u)du+\int_{(F^0_k)^{-1}(0.5)}^\infty f^1_k(v)dv\\
  =\,&
  \ln p\int_{\log_p2}^\infty\exp[\ell_k\{(F^0_k)^{-1}(p^{-a})\}-a\ln p]da\,+\\
  &
  \ln p\int_{\log_p2}^\infty\exp[\ell_k\{(F^0_k)^{-1}(1-p^{-a})\}-a\ln p]da,
\end{align*}
which means that
\begin{align*}
  &F^-(x,M)=\exp[\ell_k\{(F^0_k)^{-1}(p^{-a})\}-aM]<\infty,\\
  &F^+(x,M)=\exp[\ell_k\{(F^0_k)^{-1}(1-p^{-a})\}-aM]<\infty
\end{align*}
for all $a\in[\log_p2,\infty)$, with $M=\ln p$. Applying Lemma~\ref{l:ess} to $F^-(x,M)$ and $F^+(x,M)$ leads to the desired conclusion.

\begin{lemma}
  \label{l:supmin}
  For any function $f(x)$ and constants $c_1$ and $c_2$,
  \[
  \{\sup_xf(x)\}\wedge[\sup_x\{c_1f(x)+c_2\}]=
  \sup_x[f(x)\wedge\{c_1f(x)+c_2\}].
  \]
\end{lemma}
\textit{Proof}. First it is clear that
\[
\{\sup_xf(x)\}\wedge[\sup_x\{c_1f(x)+c_2\}]\geq\sup_x[f(x)\wedge\{c_1f(x)+c_2\}].
\]
Now fix $\epsilon>0$. By the definition of the supremum, there exist $x_1$ and $x_2$ such that
\[
f(x_1)>\sup_xf(x)-\epsilon,
\quad
c_1f(x_2)+c_2>c_1\sup_xf(x)+c_2-\epsilon.
\]
Complete the proof by defining $x^\star$ equal either $x_1$ or $x_2$ such that $f(x^\star)\geq f(x_1)\vee f(x_2)$. Then
\[
f(x^\star)\wedge\{c_1f(x^\star)+c_2\}
\geq
f(x_1)\wedge\{c_2f(x_2)+c_2\}
\geq
\{\sup_xf(x)-\epsilon\}\wedge\{c_1\sup_xf(x)+c_2-\epsilon\}.
\]

\begin{lemma}
  \label{l:R}
  Let
  \begin{equation}
    \label{eq:Wtilde}
    \widetilde{W}(t_1,t_2)
    =
    p^{1/2}\frac{\hat{S}_{12}(t_1,t_2)-S_1(t_1)S_2(t_2)}{\{\hat{S}_1(t_1)\hat{S}_2(t_2)-\hat{S}_1^2(t_1)\hat{S}_2^2(t_2)\}^{1/2}},
  \end{equation}
  where the marginal survival functions $S_k$ are known in the numerator and estimated in the denominator of $\widetilde{W}$. Then the proposed $\widehat{\mathcal{D}}$~\eqref{eq:Dhat} obeys
  \[
  R
  \equiv
  \left\vert
  \widehat{\mathcal{D}}
  -
  \sup_{\mathcal{S}}\vert\widetilde{W}\vert\right\vert
  \leq
  3(\ln\ln p)^2
  \]
  with probability approaching 1 under both $H_0$ and $H_A$ of~\eqref{eq:test}, where the set $\mathcal{S}$ is defined in~\eqref{eq:S}.
\end{lemma}
\textit{Proof}.
First upper-bound $R$ by
\begin{align*}
  R
  \leq&\,
  \sup_{\mathcal{S}}\left\vert\frac{p^{1/2}\hat{S}_{12}(t_1,t_2)-\hat{S}_1(t_1)\hat{S}_2(t_2)}{\{\hat{S}_1(t_1)\hat{S}_2(t_2)-\hat{S}_1^2(t_1)\hat{S}_2^2(t_2)\}^{1/2}}-\widetilde{W}\right\vert\\
  \leq\,&
  \sup_{\mathcal{S}}
  \frac{p^{1/2}\hat{S}_1\vert\hat{S}_2-S_2\vert}{(\hat{S}_1\hat{S}_2-\hat{S}_1^2\hat{S}_2^2)^{1/2}}
  +
  \sup_{\mathcal{S}}
  \frac{p^{1/2}S_2\vert\hat{S}_1-S_1\vert}{(\hat{S}_1\hat{S}_2-\hat{S}_1^2\hat{S}_2^2)^{1/2}}
  \equiv
  A+B.
\end{align*}
But
\[
A
=
\sup_{\mathcal{S}}
\frac{p^{1/2}\vert\hat{S}_2-S_2\vert}{(\hat{S}_2/\hat{S}_1-\hat{S}_2^2)^{1/2}}
\leq
\sup_{\mathcal{S}}
\frac{p^{1/2}\vert\hat{S}_2-S_2\vert}{(\hat{S}_2-\hat{S}_2^2)^{1/2}}
=
O_P\{(2\ln\ln p)^{1/2}\},
\]
which follows from the behavior of the studentized uniform empirical process \citep{eicker1979asymptotic,jaeschke1979asymptotic}. Similarly,
\[
B
\leq
\sup_{\mathcal{S}}
\frac{p^{1/2}\hat{S}_2\vert\hat{S}_1-S_1\vert}{(\hat{S}_1\hat{S}_2-\hat{S}_1^2\hat{S}_2^2)^{1/2}}
\sup_{\mathcal{S}}
\frac{S_2}{\hat{S}_2}
\leq
O_p\{(2\ln\ln p)^{1/2}\}\sup_{T_{2(1)}\leq t_2\leq T_{2(p)}}\left\vert\frac{S_2}{\hat{S}_2}\right\vert,
\]
where $T_{k(j)}$ is the $j$th order statistic of the $T_{kj}$. Corollary~10.5.2 of \citet{shorack1986empirical} implies that
\[
\Pr(\sup_{\infty< t_2\leq T_{2(p)}}\vert S_2/\hat{S}_2\vert\leq\ln\ln p)\rightarrow1.
\]
Therefore $\Pr\{R>3(\ln\ln p)^2\}\leq\Pr\{A+B>3(\ln\ln p)^2\}\rightarrow0$ under both $H_0$ and $H_A$.

\begin{lemma}
  \label{l:detectable}
  Consider the test
  \begin{equation}
    \label{eq:oracle_test}
    \mbox{reject $H_0$ of~\eqref{eq:test} if }\mathcal{D}>\ln p
  \end{equation}
  based on the oracle statistic $\mathcal{D}$~\eqref{eq:oracle}. Suppose $F^0_k\ne F^1_k,k=1,2$ and define
  \[
  v^-_k(x)=\esssup_{a\geq x}\{\alpha^-_k(a)-a\},
  \quad
  v^+_k(x)=\esssup_{a\geq x}\{\alpha^+_k(a)-a\}.
  \]
  Under calibration~\eqref{eq:calibrate} and Assumption~\ref{a:tails}, the sum of the type I and II errors of test~\eqref{eq:oracle_test} goes to 0 if any of the inequalities~\eqref{eq:Q1}--\eqref{eq:Q4} from Theorem~\ref{thm:adaptive} are true.
\end{lemma}
\textit{Proof}.
Since $P_{H_0}(\mathcal{D}>\ln p)=o(1)$ by Lemma~\ref{l:lil}, it remains to show that $\Pr_{H_A}(\mathcal{D}\leq\ln p)$ is also $o(1)$. For $t_k\in(0,1)$, define
\begin{equation}
  \label{eq:W}
  W(t_1,t_2)
  =
  p^{1/2}\frac{\hat{S}_{12}(t_1,t_2)-S_1(t_1)S_2(t_2)}{\{S_1(t_1)S_2(t_2)-S_1^2(t_1)S_2^2(t_2)\}^{1/2}}.
\end{equation}
Then for any $(t_1,t_2)$,
\[
\Pr_{H_A}(\mathcal{D}\leq\ln p)
\leq
\Pr_{H_A}\{\vert W(t_1,t_2)\vert\leq\ln p\}.
\]
By the triangle inequality and Chebyshev's inequality,
\begin{align*}
  \Pr_{H_A}\{\vert W(t_1,t_2)\vert\leq\ln p\}
  \leq\,&
  \Pr_{H_A}\{\vert W(t_1,t_2)-\E_{H_A}W\vert\geq\vert\E_{H_A}W(t_1,t_2)\vert-\ln p\}\\
  \leq\,&
  \frac{\var_{H_A}W(t_1,t_2)}{\{\vert\E_{H_A}W(t_1,t_2)\vert-\ln p\}^2}.
\end{align*}
Under model~\eqref{eq:mixture} the true bivariate survival function is
\begin{align*}
  S_{12}(t_1,t_2)
  =\,&
  \pi_{00}S^0_1(t_1)S^0_2(t_2)+
  \pi_{01}S^0_1(t_1)S^1_2(t_2)+
  \pi_{11}S^1_1(t_1)S^0_2(t_2)+
  \pi_{11}S^1_1(t_1)S^1_2(t_2)\\
  =\,&
  S_1(t_1)S_2(t_2)+p^{-\beta}(S^1_1-S^0_1)(S^1_2-S^0_2),
\end{align*}
where $S^0_k=1-F^0_k$, $S^1_k=1-F^1_k$, and $\pi_{ab}=\Pr(I_{1j}=a,I_{2j}=b)$. Then the expectation and variance of $W$ obey
\begin{align}
  E_{H_A}W
  =&\,
  \frac{p^{1/2-\beta}(S^1_1-S^0_1)(S^1_2-S^0_2)}
       {(S_1S_2-S_1^2S_2^2)^{1/2}},\label{eq:E}\\
  \var_{H_A}W
  =&\,
  \frac{S_{12}-S_{12}^2}{S_1S_2-S_1^2S_2^2}
  =
  (E_{H_A}W)^2
  \frac{S_{12}(1-S_{12})}{\{p^{1/2-\beta}(S^1_1-S^0_1)(S^1_2-S^0_2)\}^2}\label{eq:V}.
\end{align}
The desired result follows if there exists a $(t_1,t_2)$ such that
\begin{align}
  &\ln p/\vert\E_{H_A}W(t_1,t_2)\vert\rightarrow0,\label{eq:logp/E}\\
  &\var_{H_A}W(t_1,t_2)/\vert\E_{H_A}W(t_1,t_2)\vert^2\rightarrow0.\label{eq:V/E^2}
\end{align}

Divide $\mathbb{R}^2$ into four regions
\begin{align*}
  \mathcal{Q}_1
  &=
  \{t_1,t_2:t_1\geq(F^0_1)^{-1}(0.5),t_2\geq(F^0_2)^{-1}(0.5)\},\\
  \mathcal{Q}_2
  &=
  \{t_1,t_2:t_1<(F^0_1)^{-1}(0.5),t_2\geq(F^0_2)^{-1}(0.5)\},\\
  \mathcal{Q}_3
  &=
  \{t_1,t_2:t_1\geq(F^0_1)^{-1}(0.5),t_2<(F^0_2)^{-1}(0.5)\},\\
  \mathcal{Q}_4
  &=
  \{t_1,t_2:t_1<(F^0_1)^{-1}(0.5),t_2<(F^0_2)^{-1}(0.5)\};
\end{align*}
the result follows if there exists a $(t_1,t_2)$ in any of these regions that satisfies \eqref{eq:logp/E} and \eqref{eq:V/E^2}.

In quadrant $\mathcal{Q}_1$, define $x_1$ such that $t_1=(F^0_1)^{-1}(1-p^{-x_1})$ and $x_2$ such that $t_2=(F^0_2)^{-1}(1-p^{-x_2})$. Then $x_k\geq\log_p2,k=1,2$ and by Lemma~\ref{l:cdf}, for $p$ sufficiently large and some generic constant $C_p$ that may contain factors of $\ln p$,
\begin{align*}
  &
  \vert(S^1_1-S^0_1)(S^1_2-S^0_2)\vert\\
  =\,&
  \vert[1-F^1_1\{(F^0_1)^{-1}(1-p^{-x_1})\}-p^{-x_1}][1-F^1_2\{(F^0_2)^{-1}(1-p^{-x_2})\}-p^{-x_2}]\vert\\
  =\,&
  \vert\{p^{v^+_1(x_1)+o(1)}-p^{-x_1}\}\{p^{v^+_2(x_2)+o(1)}-p^{-x_2}\}\vert\\
  =\,&
  C_pp^{v^+_1(x_1)\vee(-x_1)+v^+_2(x_2)\vee(-x_2)}.
\end{align*}
In addition,
\[
S_k
=
(1-p^{-\beta_k})p^{-x_k}
+
p^{-\beta_k}p^{v^+_k(x_k)+o(1)}
=
C_pp^{(-x_k)\vee\{-\beta_k+v^+_k(x_k)\}},
\quad
k=1,2,
\]
and since Lemma~\ref{l:v<0} implies that $v^+_k(x_k)\leq0$ for all $x_k\geq\log_p2$, $S_1S_2=o(1)$ and therefore $S_{12}=o(1)$ in $\mathcal{Q}_1$. Therefore~\eqref{eq:E} and~\eqref{eq:V} become
\begin{align*}
  \vert E_{H_A}W\vert
  =&\,
  C_p\frac{p^{1/2-\beta+(-x_1)\vee v^+_1(x_1)+(-x_2)\vee v^+_2(x_2)}}
  {p^{[(-x_1)\vee\{-\beta_1+v^+_1(x_1)\}+(-x_2)\vee\{-\beta_2+v^+_2(x_2)\}]/2}},\\
  \frac{\var_{H_A}W}{\vert E_{H_A}W\vert^2}
  =&\,
  \frac{S_{12}\{1-o(1)\}}{\{p^{1/2-\beta}(S^1_1-S^0_1)(S^1_2-S^0_2)\}^2}\\
  =\,&
  C_p\frac{S_1S_2}{\{p^{1/2-\beta}(S^1_1-S^0_1)(S^1_2-S^0_2)\}^2}
  +
  C_p\frac{p^{-\beta}(S^1_1-S^0_1)(S^1_2-S^0_2)}{\{p^{1/2-\beta}(S^1_1-S^0_1)(S^1_2-S^0_2)\}^2}\\
  =\,&
  \frac{C_p}{\vert E_{H_A}W\vert^2}
  +
  \frac{C_p}{p^{1-\beta+(-x_1)\vee v^+_1(x_1)+(-x_2)\vee v^+_2(x_2)}}.
\end{align*}
Thus~\eqref{eq:Q1} is a sufficient condition for there to exist a $(t_1,t_2)\in\mathcal{Q}_1$ such that \eqref{eq:logp/E} and \eqref{eq:V/E^2} hold, because when $x_1+x_2<1$,
\begin{align*}
  &\,
  1-\beta+(-x_1)\vee v^+_1+(-x_2)\vee v^+_2(x_2)\\
  >&\,
  \frac{1}{2}-\beta+(-x_1)\vee v^+_1+(-x_2)\vee v^+_2(x_2)+\frac{x+y}{2}\\
  \geq&\,
  \frac{1}{2}-\beta+(-x_1)\vee v^+_1(x_1)+(-x_2)\vee v^+_2(x_2)+\frac{x_1\wedge\{\beta_1-v^+_1(x_1)\}}{2}+\frac{x_2\wedge\{\beta_2-v^+_2(x_2)\}}{2}.
\end{align*}

In quadrant $\mathcal{Q}_2$, define $x_1$ such that $t_1=(F^0_1)^{-1}(p^{-x_1})$ and $x_2$ such that $t_2=(F^0_2)^{-1}(1-p^{-x_2})$. Then again $x_k\geq\log_p2,k=1,2$,
\begin{align*}
  \vert(S^1_1-S^0_1)(S^1_2-S^0_2)\vert
  =\,&
  \vert\{1-p^{v^-_1(x_1)+o(1)}-1+p^{-x_1}\}\{p^{v^+_2(x_2)+o(1)}-p^{-x_2}\}\vert\\
  =\,&
  C_pp^{(-x_1)\vee v^-_1(x_1)}p^{(-x_2)\vee v^+_2(x_2)},
\end{align*}
and
\begin{align*}
  S_1
  =\,&
  (1-p^{-\beta_1})(1-p^{-x_1})+p^{-\beta_1}(1-p^{v^-_1(x_1)+o(1)})
  =
  O(1),\\
  S_2
  =\,&
  C_pp^{v^+_2(x_2)\vee(-x_2)}
  =
  o(1).
\end{align*}
Again $S_1S_2=o(1)$ and $S_{12}=o(1)$, so~\eqref{eq:E} and~\eqref{eq:V} become
\begin{align*}
  \vert E_{H_A}W\vert
  =&\,
  C_p\frac{p^{1/2-\beta+(-x_1)\vee v^-_1(x_1)+(-x_2)\vee v^+_2(x_2)}}{p^{\{(-x_2)\vee(-\beta_2+v^+_2(x_2))\}/2}},\\
  \frac{\var_{H_A}W}{\vert E_{H_A}W\vert^2}
  =&\,
  \frac{C_p}{\vert E_{H_A}W\vert^2}+
  \frac{C_p}{p^{1-\beta+(-x_1)\vee v^-_1(x_1)+(-x_2)\vee v^+_2(x_2)}}.
\end{align*}
Thus~\eqref{eq:Q2} is a sufficient condition for there to exist a $(t_1,t_2)\in\mathcal{Q}_2$ such that \eqref{eq:logp/E} and \eqref{eq:V/E^2} hold, because when $x_2<1$,
\begin{align*}
  &\,
  1-\beta+(-x_1)\vee v^-_1(x_1)+(-x_2)\vee v^+_2(x_2)\\
  >&\,
  \frac{1}{2}-\beta+(-x_1)\vee v^-_1(x_1)+(-x_2)\vee v^+_2(x_2)+\frac{x_2}{2}\\
  \geq&\,
  \frac{1}{2}-\beta+(-x_1)\vee v^-_1(x_1)+(-x_2)\vee v^+_2(x_2)+\frac{x_2\wedge\{\beta_2-v^+_2(x_2)\}}{2}.
\end{align*}
It can be similarly be shown that \eqref{eq:Q3} is a sufficient condition for \eqref{eq:logp/E} and \eqref{eq:V/E^2} hold for some $(t_1,t_2)\in\mathcal{Q}_3$.

Finally, in quadrant $\mathcal{Q}_4$ define $x_1$ such that $t_1=(F^0_1)^{-1}(p^{-x_1})$ and $x_2$ such that $t_2=(F^0_2)^{-1}(p^{-x_2})$, so that $x_k\geq\log_p2,k=1,2$,
\begin{align*}
  &
  \vert(S^1_1-S^0_1)(S^1_2-S^0_2)\vert\\
  =\,&
  \vert\{1-p^{v^-_1(x_1)+o(1)}-1+p^{-x_1}\}\{1-p^{v^-_2(x_2)+o(1)}-1+p^{-x_2}\}\\
  =\,&
  C_pp^{(-x_1)\vee v^-_1(x_1)+(-x_2)\vee v^-_2(x_2)},
\end{align*}
and $S_k=O(1-p^{-x_k}-p^{-\beta_k}$. Now $S_1S_2=1-o(1)$ and thus $S_{12}=1-o(1)$, so~\eqref{eq:E} and~\eqref{eq:V} become
\begin{align*}
  \vert E_{H_A}W\vert
  =\,&
  C_p\frac{p^{1/2-\beta+(-x_1)\vee v^-_1(x_1)+(-x_2)\vee v^-_2(x_2)}}{p^{\{(-x_1)\vee(-\beta_1)\vee(-x_2)\vee(-\beta_2)\}/2}},\\
  \frac{\var_{H_A}W}{\vert E_{H_A}W\vert^2}
  =\,&
  \frac{\{1-o(1)\}\{1-S_{12}(t_1,t_2)\}}{\{p^{1/2-\beta}(S^1_1-S^0_1)(S^1_2-S^0_2)\}^2}
  =
  \frac{C_p}{\vert E_{H_A}W\vert^2}
  +
  \frac{C_p}{p^{1-\beta+(-x_1)\vee v^-_1(x_1)+(-x_2)\vee v^-_2(x_2)}}.
\end{align*}
Since $\beta_k\leq1$,
\begin{align*}
  &\,
  1-\beta+(-x_1)\vee v^-_1(x_1)+(-x_2)\vee v^-_2(x_2)\\
  \geq&\,
  \frac{1}{2}-\beta+(-x)\vee p_U^-+(-y)\vee p_V^-+\frac{1}{2}\\
  \geq&\,
  \frac{1}{2}-\beta+(-x)\vee p_U^-+(-y)\vee p_V^++\frac{x\wedge\beta_U\wedge y\wedge\beta_V}{2},
\end{align*}
Therefore \eqref{eq:Q4} is a sufficient condition for there to exist a $(t_1,t_2)\in\mathcal{Q}_4$ such that \eqref{eq:logp/E} and \eqref{eq:V/E^2} hold.

\subsection{Proofs of main results}
\subsection{\label{sec:pf_adaptive}Proof of Theorem~\ref{thm:adaptive}}
It must be shown that the type I error of test~\eqref{eq:adaptive_test} goes to zero, and that the type II error goes to zero if any of inequalities~\eqref{eq:Q1}--\eqref{eq:Q4} are true.

To show that the type I error goes to zero, by Lemma~\ref{l:R},
\[
\Pr_{H_0}\{(\widehat{\mathcal{D}}>\ln p(\ln\ln p)^2+3(\ln\ln p)^2\}
\leq
\Pr_{H_0}\{\sup_{\mathcal{S}}\vert\widetilde{W}\vert>\ln p(\ln\ln p)^2\}+o(1),
\]
where $\widetilde{W}$ is defined in~\eqref{eq:Wtilde} of Lemma~\ref{l:R}. Next,
\[
\frac{1-S_1S_2}{1-\hat{S}_1\hat{S}_2}
=
\frac{1-S_2}{1-\hat{S}_1\hat{S}_2}+\frac{S_2(1-S_1)}{1-\hat{S}_1\hat{S}_2}
\leq
\frac{1-S_2}{1-\hat{S}_2}+\frac{1-S_1}{1-\hat{S}_1}\frac{S_2}{\hat{S}_2},
\]
which implies that
\[
\sup_{\mathcal{S}}\left\vert\frac{S_1S_2(1-S_1S_2)}{\hat{S}_1\hat{S}_2(1-\hat{S}_1\hat{S}_2)}\right\vert
\leq
\sup_{\mathcal{S}}\left\vert\frac{S_1}{\hat{S}_1}\right\vert
\sup_{\mathcal{S}}\left\vert\frac{S_2}{\hat{S}_2}\right\vert
\left\{
\sup_{\mathcal{S}}\left\vert\frac{1-S_2}{1-\hat{S}_2}\right\vert
+
\sup_{\mathcal{S}}\left\vert\frac{1-S_1}{1-\hat{S}_1}\right\vert
\sup_{\mathcal{S}}\left\vert\frac{S_2}{\hat{S}_2}\right\vert
\right\}.
\]
Corollary~10.5.2 of \citet{shorack1986empirical} implies that
\begin{align*}
  &\Pr(\sup_{\infty<t_1\leq T_{1(p)}}\vert S_1/\hat{S}_1\vert\leq\ln\ln p)\rightarrow1,\\
  &\Pr(\sup_{T_{1(1)}\leq t_1<\infty}\vert(1-S_1)/(1-\hat{S}_1)\vert\leq\ln\ln p)\rightarrow1,
\end{align*}
so
\[
\Pr\left\{
\sup_{\mathcal{S}}
\left\vert\frac{S_1S_2(1-S_1S_2)}{\hat{S}_1\hat{S}_2(1-\hat{S}_1\hat{S}_2)}\right\vert
\leq
(\ln\ln p)^4
\right\}\rightarrow1.
\]
Then
\begin{align*}
  \Pr_{H_0}\{\sup_{\mathcal{S}}\vert\widetilde{W}\vert>\ln p(\ln\ln p)^2\}
  \leq\,&
  \Pr_{H_0}\left\{
  \mathcal{D}\sup_{\mathcal{S}}
  \left\vert\frac{S_1S_2(1-S_1S_2)}{\hat{S}_1\hat{S}_2(1-\hat{S}_1\hat{S}_2)}\right\vert^{1/2}
  >
  \ln p(\ln\ln p)^2
  \right\}\\
  \leq\,&
  \Pr_{H_0}(\mathcal{D}>\ln p)
  +
  o(1)
  =
  o(1),
\end{align*}
where the last inequality uses Lemma~\ref{l:lil}.

To identify conditions under which the type II error of test~\eqref{eq:adaptive_test} goes to zero, by Lemma~\ref{l:R}, it suffices to identify conditions such that
\[
\Pr_{H_A}\{\sup_{\mathcal{S}}\vert\widetilde{W}\vert\leq\ln p(\ln\ln p)^2+6(\ln\ln p)^2\}+o(1),
\]
where $\widetilde{W}$ is defined in Lemma~\ref{l:R}. Next, similar to above,
\[
\frac{1-\hat{S}_1\hat{S}_2}{1-S_1S_2}
=
\frac{1-\hat{S}_2}{1-S_1S_2}+\frac{\hat{S}_2(1-\hat{S}_1)}{1-S_1S_2}
\leq
\frac{1-\hat{S}_2}{1-S_2}+\frac{1-\hat{S}_1}{1-S_1}\frac{\hat{S}_2}{S_2},
\]
and Corollary~10.5.1 of \citet{shorack1986empirical} implies that
\begin{align*}
  &\Pr(\sup_{t_1}\vert\hat{S}_1/S_1\vert\leq\ln p)\rightarrow1,\\
  &\Pr(\sup_{t_1}\vert(1-\hat{S}_1)/(1-S_1)\vert\leq\ln p)\rightarrow1,
\end{align*}
so
\[
\Pr_{H_A}\left\{
\sup_{t_1,t_2}\left\vert
\frac{\hat{S}_1\hat{S}_2(1-\hat{S}_1\hat{S}_2)}{S_1S_2(1-S_1S_2)}
\right\vert
\leq
\ln^4p
\right\}
\rightarrow1.
\]
Therefore,
\begin{align*}
  &
  \Pr_{H_A}\left\{
  \sup_{\mathcal{S}}\vert\widetilde{W}\vert
  \leq
  \ln p(\ln\ln p)^2+6(\ln\ln p)^2
  \right\}\\
  \leq\,&
  \Pr_{H_A}\left\{
  \sup_{\mathcal{S}}\vert\widetilde{W}\vert
  \sup_{t_1,t_2}\left\vert\frac{\hat{S}_1\hat{S}_2(1-\hat{S}_1\hat{S}_2)}{S_1S_2(1-S_1S_2)}\right\vert^{1/2}
  \leq
  \ln^3p(\ln\ln p)^2+6\ln^2p(\ln\ln p)^2
  \right\}
  +
  o(1)\\
  \leq&\,
  \Pr_{H_A}\left\{
  \sup_{\mathcal{S}}\vert W\vert
  \leq
  \ln^3p(\ln\ln p)^2+6\ln^2p(\ln\ln p)^2
  \right\}
  +
  o(1),
\end{align*}
where $W(t_1,t_2)$ is defined in~\eqref{eq:W} such that $\mathcal{D}=\sup_{t_1,t_2}\vert W(t_1,t_2)\vert$.
  
It will be shown below that
\begin{equation}
  \label{eq:supS}
  \Pr(\mathcal{D}\leq\ln p\vee\sup_{\mathcal{S}}\vert W\vert)
  \rightarrow1.
\end{equation}
This implies that
\begin{align*}
  &
  \Pr_{H_A}\left\{
  \sup_{\mathcal{S}}\vert\widetilde{W}\vert
  \leq
  \ln p(\ln\ln p)^2+6(\ln\ln p)^2
  \right\}\\
  \leq\,&
  \Pr_{H_A}\left\{
  \ln p\vee\sup_{\mathcal{S}}\vert W\vert
  \leq
  \ln^3p(\ln\ln p)^2+6\ln^2p(\ln\ln p)^2
  \right\}+o(1)\\
  \leq\,&
  \Pr_{H_A}\left\{
  \mathcal{D}
  \leq
  \ln^3p(\ln\ln p)^2+6\ln^2p(\ln\ln p)^2
  \right\}+o(1).
\end{align*}
In the proof of Lemma~\ref{l:detectable}, it was shown that when one of~\eqref{eq:Q1}--\eqref{eq:Q4} is true,~\eqref{eq:V/E^2} holds and $\vert E_{H_A}W\vert$ grows polynomially in $p$. This means that $\{\ln^3p(\ln\ln p)^2+6\ln^2p(\ln\ln p)^2\}/\vert E_{H_A}W\vert\rightarrow0$. Therefore following the same reasoning as in Lemma~\ref{l:detectable}, that the above probability goes to zero, which gives the desired conclusion.

It remains to show~\eqref{eq:supS}. First,
\begin{align*}
  \mathcal{D}
  =
  \max\Bigg(\,&
  \sup_{\substack{T_{1(p)}<t_1<\infty,\\T_{2(p)}<t_2<\infty}}\left\vert\frac{-p^{1/2}S_1S_2}{(S_1S_2-S_1^2S_2^2)^{1/2}}\right\vert,
  \sup_{\substack{-\infty<t_1<T_{1(1)},\\T_{2(p)}<t_2<\infty}}\left\vert\frac{-p^{1/2}S_1S_2}{(S_1S_2-S_1^2S_2^2)^{1/2}}\right\vert,
  \\
  &
  \sup_{\substack{T_{1(p)}<t_1<\infty,\\-\infty<t_2<T_{2(1)}}}\left\vert\frac{-p^{1/2}S_1S_2}{(S_1S_2-S_1^2S_2^2)^{1/2}}\right\vert,
  \sup_{\substack{-\infty<t_1<T_{1(1)},\\-\infty<t_2<T_{2(1)}}}\left\vert\frac{p^{1/2}(1-S_1S_2)}{(S_1S_2-S_1^2S_2^2)^{1/2}}\right\vert,
  \sup_{\mathcal{S}}\vert W\vert
  \Bigg).
\end{align*}
Next, since the function $x/(1-x)$ is increasing in $x$,
\[
\sup_{\substack{T_{1(p)}<t_1<\infty,\\T_{2(p)}<t_2<\infty}}\left\vert\frac{-p^{1/2}S_1S_2}{(S_1S_2-S_1^2S_2^2)^{1/2}}\right\vert
=
\left\{
\frac{pS_1(T_{1(p)})S_2(T_{2(p)})}{1-S_1(T_{1(p)})S_2(T_{2(p)})}
\right\}^{1/2}.
\]
Since the $S_k(T_{kj})$ is uniformly distributed, $S_k(T_{k(p)})$ is distributed like the minimum of $p$ independent uniforms. By Exercise~2 on p. 408 of \citet{shorack1986empirical}, $\Pr(U_{(1)}\leq p^{-1}\ln\ln p)\rightarrow1$, where $U_{(j)}$ is the $j$th order statistic of $p$ uniforms. Therefore
\begin{align*}
  &
  P\left\{
  \sup_{\substack{T_{1(p)}<t_1<\infty,\\T_{2(p)}<t_2<\infty}}\left\vert\frac{-p^{1/2}S_1S_2}{(S_1S_2-S_1^2S_2^2)^{1/2}}\right\vert
  >
  \ln p
  \right\}\\
  =\,&
  P\left\{
  \frac{pS_1(T_{1(p)})S_2(T_{2(p)})}{1-S_1(T_{1(p)})S_2(T_{2(p)})}
  >
  \ln^2p
  \cap
  S_k(T_{k(p)})\leq\frac{\ln\ln p}{p},k=1,2
  \right\}
  +o(1)\\
  =\,&
  P\left\{
  \frac{p(\ln\ln p)^2}{p^2-(\ln\ln p)^2}>\ln^2p
  \right\}
  +
  o(1)
  \rightarrow0.
\end{align*}
By similar reasoning, it can be shown that
\[
\Pr
\left\{
\sup_{\substack{-\infty<t_1<T_{1(1)},\\T_{2(p)}<t_2<\infty}}\left\vert\frac{-p^{1/2}S_1S_2}{(S_1S_2-S_1^2S_2^2)^{1/2}}\right\vert
\vee
\sup_{\substack{T_{1(p)}<t_1<\infty,\\-\infty<t_2<T_{2(1)}}}\left\vert\frac{-p^{1/2}S_1S_2}{(S_1S_2-S_1^2S_2^2)^{1/2}}\right\vert
\leq
\ln p
\right\}
\rightarrow1.
\]
Finally, because $(1-x)^2/(x-x^2)$ is decreasing in $x$, $S(T_{k(1)})$ is distributed like the maximum of $p$ uniforms, and $\Pr(U_{(p)}>1-p^{-1}\ln\ln p)\rightarrow1$,
\[
\Pr
\left\{
\sup_{\substack{-\infty<t_1<T_{1(1)},\\-\infty<t_2<T_{2(1)}}}\left\vert\frac{p^{1/2}(1-S_1S_2)}{(S_1S_2-S_1^2S_2^2)^{1/2}}\right\vert,
\leq
\ln p
\right\}
\rightarrow1.
\]
Together these imply~\eqref{eq:supS}.

\subsection{\label{sec:pf_undetectable}Proof of Theorem~\ref{thm:undetectable}}
The squared Hellinger distance between two distributions $P_0$ and $P_1$, with densities $p_0$ and $p_1$ with respect to the Lebesgue measure $\mu$, is defined as
\[
H^2(P_0,P_1)=\frac{1}{2}\int(p_0^{1/2}-p_1^{1/2})^2d\mu.
\]
If $P_0$ and $P_1$ are the distributions of $(T_{1j},T_{2j})$ under $H_0$ and $H_A$ of~\eqref{eq:test}, respectively, then by Theorem~13.1.3 of \citet{lehmann2005testing}, the sum of the type I and II errors of any test goes to at least one if $pH^2(P_0,P_1)\rightarrow0$. It remains to show that conditions \eqref{eq:undetectable1} and~\eqref{eq:undetectable2} imply $H^2(P_0,P_1)=o(p^{-1})$.

For compactness of notation define the function
\[
q(t_1,t_2)=
\left\{
1-\left(
  1+
  \frac{p^{-\beta}\{L_1(t_1)-1\}\{L_2(t_2)-1\}}
       {[1+p^{-\beta_1}\{L_1(t_1)-1\}][1+p^{-\beta_2}\{L_2(t_2)-1\}]}
       \right)^{1/2}
\right\}^2,
\]
where $L_k(t_k)=f^1_k(t_k)/f^0_k(t_k),k=1,2$ are likelihood ratios. Also define the sets
\begin{align*}
  \mathcal{I}_1&=\{t_1,t_2:L_1(t_1)<1,L_2(t_2)<1\},\\
  \mathcal{I}_2&=\{t_1,t_2:1\leq L_1(t_1),L_2(t_2)<1\},\\
  \mathcal{I}_3&=\{t_1,t_2:L_1(t_1)<1,1\leq L_2(t_2)\},\\
  \mathcal{I}_4&=\{t_1,t_2:1\leq L_1(t_1),1\leq L_2(t_2)\}.
\end{align*}
By definition the $L_k$ are always positive. Then the squared Hellinger distance satisfies
\[
2H^2(P_0,P_1)
=
\sum_{r=1}^4\int_{\mathcal{I}_r}qf_1f_2dt_1dt_2,
\]
where $f_k=(1-\pi_k)f^0_k+\pi_kf^1_k$ is the marginal densities of $T_{kj}$. Each integral in this sum will be shown to be $o(p^{-1})$ under~\eqref{eq:undetectable1} and~\eqref{eq:undetectable2}.

First, on $\mathcal{I}_1$ the term inside the square root in $q(t_1,t_2)$ is maximized when $L_k=0,k=1,2$ and is always larger than one for $p>1$. Using this Lemma~4.2 of \citet{cai2014optimal}, which states that  $\{1-(1+t)^{1/2}\}^2\leq t\wedge t^2$ for $t\geq0$,
\begin{align*}
  \int_{\mathcal{I}_1}qf_1f_2dt_1dt_2
  \leq&\,
  \int_{\mathcal{I}_1}
  \left(
  1-\left[
    1+
    \frac{p^{-\beta}}
         {(1-p^{-\beta_1})(1-p^{-\beta_2})}
         \right]^{1/2}
  \right)^2
  f_1f_2dt_1dt_2\\
  \leq&\,
  \frac{p^{-\beta}}
       {(1-p^{-\beta_1})(1-p^{-\beta_2})}
  \wedge
  \frac{p^{-2\beta}}
       {(1-p^{-\beta_1})^2(1-p^{-\beta_2})^2}\\
  =&\,
  o(p^{-1}),
\end{align*}
where the last line follows because $\beta>1/2$ under weak latent dependence.

To upper-bound $q(t_1,t_2)$ on $\mathcal{I}_2$, it is easy to show that $\partial q/\partial L_2\leq0$, which implies that $q$ is maximized when $L_2=0$. Therefore
\begin{align*}
  \int_{\mathcal{I}_2}qf_1f_2dt_1dt_2
  \leq&\,
  \int_{\mathcal{I}_2}\left[1-\left\{1-\frac{p^{-\beta}}{1-p^{-\beta_2}}\frac{L_1-1}{1+p^{-\beta_1}(L_1-1)}\right\}^{1/2}\right]^2f_1f_2dt_1dt_2\\
  \leq&\,
  \frac{p^{-2\beta}}{(1-p^{-\beta_2})^2}\int_{\mathcal{I}_2}\frac{(L_1-1)^2}{\{1+p^{-\beta_1}(L_1-1)\}^2}f_1f_2dt_1dt_2\\
  =&\,
  \frac{p^{-2\beta}}{(1-p^{-\beta_2})^2}\int_{\mathcal{I}_2}\frac{(L_1-1)^2}{1+p^{-\beta_1}(L_1-1)}f^0_1f_2dt_1dt_2,
\end{align*}
where the second inequality follows the facts that $\{1-(1-x)^{1/2}\}^2<x^2$ for $x\in[0,1]$.

Next divide $\mathcal{I}_2$ into disjoint subsets
\begin{align*}
  \mathcal{I}_{21}
  &=
  \{t_1,t_2:1\leq L_1(t_1),t_1\leq(F^0_1)^{-1}(0.5),L_2(t_2)<1\},\\
  \mathcal{I}_{22}
  &=
  \{t_1,t_2:1\leq L_1(t_1),t_1>(F^0_1)^{-1}(0.5),L_2(t_2)<1\}.
\end{align*}
On $\mathcal{I}_{21}$ make the change of variables $t_1\mapsto (F^0_1)^{-1}(p^{-a}),a\geq\log_p2$, such that
\[
dt_1=-\ln p\frac{F^0_1(t_1)}{f^0_1(t_1)}da=-\frac{p^{-a}\ln p}{f^0_1\{(F^0_1)^{-1}(p^{-a})\}}da.
\]
Similarly, on $\mathcal{I}_{22}$ use $t_1\mapsto(F^0_1)^{-1}(1-p^{-a}),a>\log_p2$, which implies
\[
dt_1=\frac{p^{-a}\ln p}{f^0_1\{(F^0_1)^{-1}(p^{-a})\}}da.
\]

Finally, Assumption~\ref{a:tails} implies that for $p$ sufficiently large, there is a small $\delta>0$ such that on $a\geq\log_p2$, $L_1\{(F^0_1)^{-1}(p^{-a})\}\leq p^{\alpha_1^-(a)+\delta}$ and $L_1\{(F^0_1)^{-1}(1-p^{-a})\}\leq p^{\alpha_1^+(a)+\delta}$. Therefore for $p$ large enough and a generic constant $C_p$ that contains a $\ln p$ factor,
\begin{align*}
  \int_{\mathcal{I}_2}qf_1f_2dt_1dt_2
  \leq&\,
  C_pp^{-2\beta}\int_{\mathcal{I}_{21}}
  \frac{[L_1\{(F^0_1)^{-1}(p^{-a})\}-1]^2}{1+p^{-\beta_1}[L_1\{(F^0_1)^{-1}(p^{-a})\}-1]}p^{-a}f_2dadt_2+\,\\
  &
  C_pp^{-2\beta}\int_{\mathcal{I}_{22}}
  \frac{[L_1\{(F^0_1)^{-1}(1-p^{-a})\}-1]^2}{1+p^{-\beta_1}[L_1\{(F^0_1)^{-1}(1-p^{-a})\}-1]}p^{-a}f_2dadt_2\\
  \leq&\,
  C_pp^{-2\beta}\int_{\left\{\substack{0\leq\alpha_1^-(a)+\delta,\\a\geq\log_p2}\right\}}
  \frac{(p^{\alpha_1^-+\delta}-1)^2}{1+p^{-\beta_1}(p^{\alpha_1^-+\delta}-1)}p^{-a}da+\,\\
  &
  C_pp^{-2\beta}\int_{\left\{\substack{0\leq\alpha_1^+(a)+\delta,\\a>\log_p2}\right\}}
  \frac{(p^{\alpha_1^++\delta}-1)^2}{1+p^{-\beta_1}(p^{\alpha_1^++\delta}-1)}p^{-a}da.
\end{align*}
Since by Assumption~\ref{a:tails} the function $\alpha_1(a)=\alpha_1^-(a)\vee\alpha_1^+(a)$ is positive on a set of non-zero Lebesgue measure,
\begin{align*}
  \int_{\mathcal{I}_2}qf_1f_2dt_1dt_2
  \leq&\,
  C_pp^{-2\beta}\int_{\log_p2}\left\{
  \frac{p^{2(\alpha_1^-+\delta)}}{1+p^{-\beta_1+\alpha_1^-+\delta}}p^{-a}+
  \frac{p^{2(\alpha_U^++\delta)}}{1+p^{-\beta_1+\alpha_1^++\delta}}p^{-a}\right\}da\\
  \leq&\,
  C_pp^{-2\beta}\int_{\log_p2}
  p^{(\alpha_1+\delta)+\{(\alpha_1+\delta)\wedge\beta_1\}-a}da.
\end{align*}

Thus by Lemma~\ref{l:ess} and \eqref{eq:undetectable1},
\[
\int_{\mathcal{I}_2}qf_1f_2dt_1dt_2
\leq
C_pp^{-2\beta+\esssup_{a\geq\log_p2}[(\alpha_1+\delta)+\{(\alpha_1+\delta)\wedge\beta_1\}-a]}
=
o(p^{-1}).
\]
Similar reasoning shows that the integral over $\mathcal{I}_3$ is $o(p^{-1})$ as well.

To complete the proof, divide the fourth region $\mathcal{I}_4$ into disjoint subsets
\begin{align*}
  \mathcal{I}_{41}
  &=
  \{t_1,t_2:1\leq L_1(t_1),t_1\leq(F^0_1)^{-1}(0.5),1\leq L_2(t_2),t_2\leq(F^0_2)^{-1}(0.5)\},\\
  \mathcal{I}_{42}
  &=
  \{t_1,t_2:1\leq L_1(t_1),t_1>(F^0_1)^{-1}(0.5),1\leq L_2(t_2),t_2\leq(F^0_2)^{-1}(0.5)\},\\
  \mathcal{I}_{43}
  &=
  \{t_1,t_2:1\leq L_1(t_1),t_1\leq(F^0_1)^{-1}(0.5),1\leq L_2(t_2),t_2>(F^0_2)^{-1}(0.5)\},\\
  \mathcal{I}_{44}
  &=
  \{t_1,t_2:1\leq L_1(t_1),t_1>(F^0_1)^{-1}(0.5),1\leq L_2(t_2),t_2>(F^0_2)^{-1}(0.5)\}.
\end{align*}
On $\mathcal{I}_{41}$ let $t_1\mapsto(F^0_1)^{-1}(p^{-a_1})$ and $t_2\mapsto(F^0_2)^{-1}(p^{-a_2})$. Since by Assumption~\ref{a:tails} the function $\alpha_2(a)=\alpha_2^-(a)\vee\alpha_2^+(a)$ is positive on a set of non-zero Lebesgue measure,
\begin{align*}
  \int_{\mathcal{I}_{41}}qf_1f_2dt_1dt_2
  \leq
  C_p\int_{\{a_1,a_2\geq\log_p2\}}\,&
  \left[1-\left\{1+
    \frac{p^{-\beta+\alpha_1^-+\alpha_2^-+2\delta}}
         {(1+p^{-\beta_1+\alpha_1^-+\delta})(1+p^{-\beta_2+\alpha_2^-+\delta})}
         \right\}^{1/2}\right]^2\\
  &\,
  (1+p^{-\beta_1+\alpha_1^-+\delta})(1+p^{-\beta_2+\alpha_1^-+\delta})p^{-a_1-a_2}da_1da_2\\
  \leq
  C_p\int_{\{a_1,a_2\geq\log_p2\}}\,&
  \left\{p^{-\beta+\alpha_1^-+\alpha_2^-+2\delta}
  \wedge
  \frac{p^{-2\beta+2\alpha_1^-+2\alpha_2^-+4\delta}}
       {(1+p^{-\beta_1+\alpha_1^-+\delta})(1+p^{-\beta_2+\alpha_2^-+\delta})}\right\}\\
  &\,
  p^{-a_1-a_2}da_ada_2,
\end{align*}
because $\{1-(1+t)^{1/2}\}^2\leq t\wedge t^2$ by Lemma~4.2 of \citet{cai2014optimal}. This in turn is bounded by
\[
C_p\int_{\{a_1,a_2\geq\log_p2\}}
[p^{-\beta+\alpha_1^-+\alpha_2^-+2\delta}
  \wedge
  p^{-2\beta+\alpha_1^-+\alpha_2^-+2\delta+\{(\alpha_1^-+\delta)\wedge\beta_1\}+\{(\alpha_2^-+\delta)\wedge\beta_2\}}]
p^{-a_1-a_2}da_1da_2.
\]
Corresponding calculations over the other three subsets of $\mathcal{I}_4$ imply that the integral of $q$ over this region is at most 
\[
C_pp^{\esssup_{a_1,a_2\geq\log_p2}(\{-\beta+\alpha_1+\alpha_2+2\delta\}\wedge[-2\beta+\alpha_1+\alpha_2+2\delta+\{(\alpha_1+\delta)\wedge\beta_2\}+\{(\alpha_2+\delta)\wedge\beta_2\}]-a_1-a_2)},
\]
which is $o(p^{-1})$ when \eqref{eq:undetectable2} holds.

\subsection{\label{sec:pf_boundary}Proof of Theorem~\ref{thm:boundary}}
By Assumption~\ref{a:sto_ord}, $F^1_k\{(F^0_k)^{-1}(p^{-x})\}\leq p^{-x}$ and $F^1_k\{(F^0_k)^{-1}(1-p^{-x})\}\leq1-p^{-x}$. Combining this with Lemma~\ref{l:cdf} implies
\[
p^{v^-_k(x)+o(1)}\leq p^{-x}
=
1-(1-p^{-x})
\leq
1-\{1-p^{v^+_k(x)+o(1)}\}.
\]
These inequalities lead to several useful facts when $x\geq\log_p2$: for $k=1,2$,
\begin{align}
  v^+_k(x)
  \geq\,&
  -x
  \geq
  v^-_k(x),\label{eq:fact1}\\
  v^+_k(x)
  =\,&
  v^+_k(x)\vee v^-_k(x)
  =
  \esssup_{a\geq x}\{\alpha^+_k(a)-a\}\vee\esssup_{a\geq x}\{\alpha^-_k(a)-a\}\nonumber\\
  =\,&
  \esssup_{a\geq x}\{\alpha_k(a)-a\},\label{eq:fact2}
\end{align}
with $\alpha_k(a)=\alpha^+_k(a)\vee\alpha^-_k(a)$ as defined as in Assumption~\ref{a:tails}.

It must be shown that the interior of the complement of the undetectable region, defined by Theorem~\eqref{eq:undetectable1}, equals the detectable region, defined by Theorem~\ref{thm:adaptive}. No test can have a sum of type I and II errors less than one in the undetectable region, which implies that the interior of its complement contains the detectable region. It remains to show that this interior is also a subset of the detectable region, in other words, that at least one of the detectable region inequalities~\eqref{eq:Q1}--\eqref{eq:Q4} is implied when one of the undetectable region inequalities~\eqref{eq:undetectable1}--\eqref{eq:undetectable2} is false.

It helps to re-express~\eqref{eq:Q1}--\eqref{eq:Q4} when the stochastic ordering of Assumption~\ref{a:sto_ord} holds. First, by Proposition~3.5 of \citet{phu1996essential}, the supremum and essential supremum with respect to the Lebesgue measure are equal for lower semi-continuous functions. Next, by~\eqref{eq:fact1} and~\eqref{eq:fact2}, \eqref{eq:Q1} becomes
\[
0
<
\frac{1}{2}-\beta
+
\esssup_{\substack{x_1,x_2>0,\\x_1+x_2<1}}
\left[
  \sum_{k=1}^2
  \esssup_{a\geq x_k}\left\{\alpha_k(a)-a+\frac{x_k}{2}\right\}
  \wedge
  \esssup_{a\geq x_k}\left\{\frac{\alpha_k(a_k)-a_k+\beta_k}{2}\right\}
  \right].
\]
Using Lemma~\ref{l:supmin}, the essential supremum above equals
\begin{align*}
  \esssup_{\substack{x_1,x_2>0,x_1+x_2<1,\\a_1\geq x_1,a_2\geq x_2}}\Bigg(
  &\left[\sum_{k=1}^2\left\{\alpha_k(a_k)-a_k+\frac{x_k}{2}\right\}\right]
    \wedge
    \left\{\alpha_1(a_1)-a_1+\frac{x_1}{2}+\frac{\alpha_2(a_2)-a_2+\beta_2}{2}\right\}\wedge\\
    &\left\{\frac{\alpha_1(a_1)-a_1+\beta_1}{2}+\alpha_2(a_2)-a_2+\frac{x_2}{2}\right\}\wedge
    \left\{\sum_{k=1}^2\frac{\alpha_k(a_k)-a_k+\beta_k}{2}\right\}
    \Bigg),
\end{align*}
so using Lemma~\ref{l:supmin} again and taking the suprema with respect to the $x_k$ means that~\eqref{eq:Q1} is equivalent to
\begin{equation}
  \label{eq:Q1_reexpress}
  \begin{aligned}
    0
    <
    \frac{1}{2}-\beta
    +
    \esssup_{a_1,a_2>0}\bigg[
      &\left\{\alpha_1(a_1)-a_1+\alpha_2(a_2)-a_2+\frac{(a_1+a_2)\wedge1}{2}\right\}\wedge\\
      &\left\{\alpha_1(a_1)-a_1+\frac{a_1\wedge1}{2}+\frac{\alpha_2(a_2)-a_2+\beta_2}{2}\right\}\wedge\\
      &\left\{\frac{\alpha_1(a_1)-a_1+\beta_2}{2}+\alpha_2(a_2)-a_2+\frac{a_2\wedge1}{2}\right\}\wedge\\
      &\left\{\frac{\alpha_1(a_1)-a_1+\beta_1}{2}+\frac{\alpha_2(a_2)-a_2+\beta_2}{2}\right\}\bigg].
  \end{aligned}
\end{equation}
By similar reasoning,~\eqref{eq:Q2} and~\eqref{eq:Q3} are equivalent to
\begin{equation}
  \label{eq:Q2,3_reexpress}
  0
  <
  \frac{1}{2}-\beta
  +
  \esssup_{a>0}\left[
    \left\{\alpha_k(a)-a+\frac{a\wedge1}{2}\right\}
    \wedge
    \left\{\frac{\alpha_k(a)-a+\beta_k}{2}\right\}
    \right],
  \quad
  k=1,2.
\end{equation}
Finally, by~\eqref{eq:fact1} inequality~\eqref{eq:Q4} becomes
\[
0
<
\esssup_{x_1,x_2>0}
\left(
\frac{1}{2}-\beta
-x_1-x_2+
\frac{x_1\wedge\beta_1\wedge x_2\wedge\beta_2}{2}
\right)
=
\frac{1}{2}-\beta.
\]
Since by assumption $\beta>1/2$ from calibration~\eqref{eq:calibrate}, it must only be shown that either~\eqref{eq:Q1_reexpress} or~\eqref{eq:Q2,3_reexpress} holds when either~\eqref{eq:undetectable1} or~\eqref{eq:undetectable2} are false.

Now suppose~\eqref{eq:undetectable1} is false. Then there exists an $a^\star>0$ such that
\[
0
<
1-2\beta+\alpha_k(a^\star)+\alpha_k(a^\star)\wedge\beta_k-a^\star.
\]
For simplicity let $\alpha_k^\star$ denote $\alpha_k(a_k^\star)$. Then the previous inequality implies
\begin{align*}
  0
  <\,&
  \frac{1}{2}-\beta+\left(\alpha_k^\star-\frac{a^\star}{2}\right)
  \wedge
  \frac{\alpha_k^\star+\beta_k-a^\star}{2},\\
  0
  <\,&
  1-\beta+\alpha_k^\star-a_k^\star+(\beta_k-\beta)
  <
  \frac{1}{2}-\beta+\alpha_k^\star-a_k^\star+\frac{1}{2}.
\end{align*}
since $\beta>\beta_1\vee\beta_2$ from~\eqref{eq:calibrate}. Therefore~\eqref{eq:Q2,3_reexpress} holds when~\eqref{eq:undetectable1} is false.

Now suppose~\eqref{eq:undetectable2} is false. Then there exists $a_1^\star,a_2^\star>0$ such that
\begin{equation}
  \label{eq:un2_false}
  \begin{aligned}
    0
    <\,&
    1-\beta+\alpha_1(a_1^\star)+\alpha_2(a_2^\star)-a_1^\star-a_2^\star,\\
    0
    <\,&
    1-2\beta+\alpha_1(a_1^\star)+\alpha_2(a_2^\star)+\alpha_1(a_1^\star)\wedge\beta_1+\alpha_2(a_2^\star)\wedge\beta_2-a_1^\star-a_2^\star.
  \end{aligned}
\end{equation}
These inequalities imply
\begin{align*}
  0
  <\,&
  1-\beta+\alpha_1^\star+\alpha_2^\star-a_1^\star-a_2^\star
  =
  \frac{1}{2}-\beta+\alpha_1^\star+\alpha_2^\star+\frac{1}{2},\\
  0
  <\,&
  1-2\beta+2\alpha_1^\star+2\alpha_2^\star-a_1^\star-a_2^\star,\\
  0
  <\,&
  1-2\beta+\alpha_1^\star+\alpha_2^\star+\beta_1+\beta_2-a_1^\star-a_2^\star,
\end{align*}
which correspond to the first and fourth terms inside the essential supremum of~\eqref{eq:Q1_reexpress}. They also imply
\begin{align*}
  0
  <\,&
  1-2\beta+2\alpha_1^\star-a_1^\star+\alpha_2^\star+\beta_2-a_2^\star,\\
  0
  <\,&
  1-\beta+\alpha_1^\star-a_1^\star+\alpha_2^\star-a_2^\star
  <
  \frac{1}{2}-\beta+\alpha_1^\star-a_1^\star+\frac{1}{2}+\frac{\alpha_2^\star-a_2^\star+\beta_2}{2},
\end{align*}
which correspond to the second term inside the essential supremum of~\eqref{eq:Q1_reexpress}. The last inequality above follows because
\[
\alpha_2^\star-a_2^\star-\frac{\alpha_2^\star-a_2^\star+\beta_2}{2}
=
\frac{\alpha_2^\star-a_2^\star-\beta_2}{2}
\leq
0
\]
by Lemma~\ref{l:v<0}. It can be similarly shown that~\eqref{eq:un2_false} imply the third term inside the essential supremum of~\eqref{eq:Q1_reexpress} as well. Therefore~\eqref{eq:Q1_reexpress} holds when~\eqref{eq:undetectable2} is false.

\end{document}